\documentclass[journal]{IEEEtran}

\usepackage{times,graphics,amssymb,amsmath,bm,cite}
\usepackage[dvips]{graphicx}
\usepackage[active]{srcltx}  
\usepackage{epsfig,psfig,latexsym}

\newcommand{\sgn}{\mbox{sgn}}
\newcommand{\eeq}{\end{equation}}

\newcommand{\br}{\mbox{\boldmath $r$}}

\newcommand{\bs}{\mbox{\boldmath $s$}}

\newcommand{\bb}{\mbox{\boldmath $b$}}

\newcommand{\bM}{\mbox{\boldmath $M$}}

\newcommand{\bP}{\mbox{\boldmath $P$}}

\newcommand{\bH}{\mbox{\boldmath $H$}}

\newcommand{\bc}{\mbox{\boldmath $c$}}
\newcommand{\bS}{\mbox{\boldmath $S$}}

\newcommand{\bU}{\mbox{\boldmath $U$}}

\newcommand{\bLambda}{{\bf \Lambda}}

\newcommand{\bu}{\mbox{\boldmath $u$}}

\newcommand{\bn}{\mbox{\boldmath $n$}}

\newcommand{\bd}{\mbox{\boldmath $d$}}

\newcommand{\bI}{\mbox{\boldmath $I$}}

\newcommand{\bD}{\mbox{\boldmath $D$}}

\newcommand{\ds}{\displaystyle}

\newcommand{\K}{{\cal K}}
\newcommand{\G}{{\cal G}}
\newcommand{\N}{{\cal N}_0}
\def\R{{\cal R}}
\def\S{{\cal S}}

\newcommand{\beq}{\begin{equation}}

\newcounter{MYtempeqncnt}

\newcommand{\gammabar}{\bar{\gamma}}
\newcommand{\Prob}{\mbox{Prob}}

\begin{document}


\title{Joint Receiver and Transmitter Optimization for Energy-Efficient CDMA Communications}
\author{{Stefano Buzzi, {\em Senior Member, IEEE},  and H. Vincent Poor, {\em Fellow, IEEE}}\\
\thanks{Stefano Buzzi is with DAEIMI, University of Cassino, Via G. Di Biasio, 43, I-03043 Cassino (FR), Italy (e-mail: buzzi@unicas.it);
H. Vincent Poor is with the School of Engineering and Applied Science, Princeton University, Princeton, NJ, 08544, USA (e-mail: poor@princeton.edu). \newline
This paper was partly presented at the {\em 2007 European Wireless Conference}, Paris, France, April 2007, and at the {\em 2007 IEEE International Symposium on Information Theory}, Nice, France, June 2007.
This research was supported in part by the U. S. Air Force Research Laboratory under Cooperative Agreement No.
FA8750-06-1-0252 and in part by the U. S. Defense Advanced Research Projects Agency under Grant HR0011-06-1-0052.
}
} \maketitle
\date{May 20, 2007}

\markboth{To appear on IEEE JSAC - special issue on Multiuser Detection
for Advanced Communication Systems and Networks}{SB \& HVP: Joint transceiver optimization for energy-efficient CDMA communications}


\begin{abstract}
This paper focuses on the cross-layer issue of joint multiuser detection and resource allocation for energy efficiency in wireless CDMA networks. In particular, assuming that a linear multiuser detector is adopted in the uplink receiver, the case considered is that in which each terminal is allowed to vary its transmit power, spreading code, and uplink receiver in order to maximize its own utility, which is defined as the ratio of data throughput to transmit power. Resorting to a game-theoretic formulation, a non-cooperative game for utility maximization is formulated, and it is proved that a unique Nash equilibrium exists, which, under certain conditions, is also Pareto-optimal.
Theoretical results concerning the relationship between the problems of SINR maximization and MSE minimization are given, and, resorting to the tools of large system analysis, a new distributed power control algorithm is implemented, based on very little prior information about the user of interest. The utility profile achieved by the active users in a large CDMA system is also computed, and, moreover, the centralized socially optimum solution is analyzed.
Considerations on the extension of the proposed framework to a multi-cell scenario are also briefly detailed.
Simulation results confirm that the proposed non-cooperative game largely outperforms competing alternatives, and that it exhibits a quite small performance loss with respect to the socially optimum solution, and only in the case in which the users number exceeds the processing gain. Finally, results also show an excellent agreement between the theoretical closed-form formulas based on large system analysis and the outcome of numerical experiments.


\end{abstract}
%

\section{Introduction}

\IEEEPARstart{A}{ substantial} amount of research has been carried out on multiuser detection for code division multiple access (CDMA) networks over the last thirty years. Starting from the pioneering work of Verd\'u, who derived the optimal, minimum error probability, multiuser receiver along with fundamental suboptimal multiuser detectors such as the decorrelating receiver, significant progress has been made over the years. Several relevant issues such as approximate implementations of the optimal multiuser detector, the impact of fading on multiuser detection structures, the synthesis of adaptive, possibly blind, multiuser detection algorithms, and the joint multiuser detection and channel equalization problem, have been tackled and thoroughly investigated. Results regarding these research issues are surveyed, among the others, in the textbooks \cite{verdu,wangpoor}. In the recent past, a new trend has emerged, i.e. the so-called cross-layer approach. Roughly speaking, the basic idea here is to perform joint optimization of procedures that are implemented in different layers of the network protocol stack, so as to outperform solutions based on single optimization of the procedures of each network layer. Regarding CDMA systems, the cross layer approach has mainly focused on the problem of integrating physical layer issues, such as multiuser detection and channel estimation, with network level issues, such as call admission control, power control, and, more generally, resource allocation \cite{crosslayer}. In keeping with this recent trend, this paper focuses on the issue of joint multiuser detection and resource allocation in order to achieve energy efficiency in wireless CDMA networks. The results of this paper are mainly based on two powerful mathematical tools, namely \emph{game theory} and \emph{large system analysis}.

Game theory \cite{gtbook} is a branch of mathematics that has been applied primarily in economics and other social sciences to study the interactions among several autonomous subjects with contrasting interests. More recently, it has been discovered that it can also be used for the design and analysis of communication systems, mostly with application to resource allocation algorithms \cite{gt}, and, in particular, to power control \cite{yates}.
As examples, the reader is referred to \cite{nara1,nara2,SaraydarPhD}. In these papers, for a multiple access wireless data network,
noncooperative and cooperative games are introduced, wherein each user chooses its  transmit power in order to maximize its own utility, defined as the ratio of the throughput to transmit power.
While the above papers consider the issue of power control assuming that a conventional matched filter is available at the receiver, the recent paper \cite{meshkati} considers the cross-layer problem of joint linear receiver design and power control so as to maximize the utility of each user: it is thus shown in \cite{meshkati} that the inclusion of receiver design in the considered game brings remarkable advantages. This same utility function is also used in \cite{bacci} for energy-efficient power control in ultra-wideband (UWB) communications, while the survey paper \cite{meshkati2} reviews recent advances in the application of a game-theoretic framework for energy-efficient resource allocation.

Large system analysis (LSA) is a relatively new mathematical tool, first introduced in \cite{tse},  that has recently emerged in the analysis of CDMA systems.  In summary, \cite{tse} has revealed that
in a CDMA system with processing gain and number of users both increasing without bound but with their ratio fixed,
and with randomly chosen, unit-norm, spreading codes, the Signal-to-Interference plus Noise Ratio (SINR) of each user for the case in which a linear minimum mean square error (MMSE) receiver is adopted converges in probability to a non-random constant.
In particular, denoting by $K$ the number of active users, by $N$ the system processing gain, by $\N/2$ the additive thermal noise power spectral density (PSD) level, and by $E_P [ \cdot]$ the expectation with respect to the limiting empirical distribution $F$ of the received powers of the interferers, the SINR of the MMSE receiver for the $k$-th user, say $\gamma_k$, converges, for $K, N \rightarrow \infty$, $K/N=\alpha=\mbox{constant}$, in probability to $\gamma_k^*$ the unique solution of the equation
\beq
\gamma_k^*= \ds \frac{P_k}{\N/2 + \alpha E_P \left[
\frac{P P_k}{P_k +P \gamma_k^*}
\right]} \; ,
\label{eq:HT}
\eeq
with $P_k$ the received power for the $k$-th user. Interestingly, the limiting SINR depends only on the limiting empirical distribution of the received powers of the interferers,  the load $\alpha$, the thermal noise level and the received power of the user of interest, while being independent of the actual realization of the received powers of the interferers and of the spreading codes of the active users. LSA is now a well-established mathematical tool  for design and analysis of communication systems (see, e.g., \cite{evans1,evans2,mimo}, to cite a few).

\subsection{Summary of the results}
This paper is the first in this area that considers the cross-layer issue of utility maximization with respect to the choice of linear multiuser detector, spreading code and transmit power. Using game theory and LSA, the following contributions are given here.

\begin{itemize}
\item[-]
We generalize the non-cooperative game considered in \cite{meshkati} by considering utility maximization with respect to the linear uplink multiuser receiver, transmit power and spreading code assignment. We will show that the newly considered non-cooperative game admits a unique Nash equilibrium, which, for the case in which the number of users does not exceed the system processing gain, is also Pareto-optimal.
\item[-]
As an introductory step to  the previous item, we also formulate a non-cooperative game for SINR maximization with respect to linear multiuser detector and spreading code choice, and show that this game admits a unique Nash equilibrium point that is also Pareto-optimal.
\item[-]
Using LSA, we design a
new distributed power control algorithm that needs very little prior information (i.e. the channel gain for the user of interest) to be implemented. This algorithm may be integrated in the utility maximizing non-cooperative game of \cite{meshkati}.
\item[-]
Using LSA, we are able to predict the utility and SINR profile across users in a large CDMA system, for both the cases in which spreading code optimization is either considered or not considered.
\item[-]
Using LSA, we are able to derive the socially optimum solution to the problem of utility maximization with equal SINR constraint; numerical results will show that the performance loss incurred by the proposed non-cooperative game with respect to the socially optimum solution is quite negligible.
\item[-]
We also consider the extension of our framework to a multi-cell scenario. In particular, we consider the issue of non-cooperative utility maximization in a multi-cell system with predetermined base station assignment, and show that, as the number of users does not exceeds the processing gain, this game admits a unique Nash equilibrium which is also Pareto-optimal.
\end{itemize}

\subsection{Outline of the paper}
The rest of this paper is organized as follows. The next section contains some preliminaries on the considered payoff function and on the system model of interest. Section III dwells on the definition of Nash equilibrium and Pareto optimality, and, also, provides an interesting result on the equilibrium point of the SINR-maximizing non-cooperative game. In section IV the non-cooperative game for utility maximization with respect to the choice of linear multiuser detection, power control and spreading code is described and analyzed. LSA is used in Section V to derive a distributed power control procedure that can be implemented based on little prior information; it is shown that this
algorithm may be used to obtain distributed implementations of the non-cooperative game proposed in \cite{meshkati}.
In Section VI, LSA is used in order to predict the SINR and utility profile across users in a large CDMA system, while Section VII contains the discussion on the socially optimum, equal SINR, cooperative game.
Section VIII considers the extension of the considered non-cooperative games to a multi-cell scenario, wherein out-of-cell interference is properly taken into account.
Finally, numerical results are illustrated in Section IX, while Section X contains the eventual wrap-up of the paper.

\section{System model and problem statement}
Consider the uplink of a $K$-user synchronous, single-cell, direct-sequence code division multiple access (DS/CDMA) network with processing gain $N$ and subject to flat fading. After chip-matched filtering and sampling at the chip-rate, the $N$-dimensional received data vector, say $\br$, corresponding to one symbol interval, can be written as
\beq
\br=\ds \sum_{k=1}^{K}\sqrt{p_k} h_k b_k \bs_k + \bn \; ,
\label{eq:r}
\eeq
wherein $p_k$ is the transmit power of the $k$-th user\footnote{To simplify subsequent notation, we assume that the transmitted power $p_k$ subsumes also the gain of the transmit and receive antennas.}, $b_k\in \{-1,1\}$ is the information symbol of the $k$-th user, and $h_k$ is the real\footnote{We assume here, for simplicity, a real channel model; generalization to practical channels, with I and Q components, is straightforward.} channel gain between the $k$-th user's transmitter and the access point (AP); the actual value of $h_k$ depends on both the distance of the $k$-th user's terminal from the AP and the channel fading fluctuations. The $N$-dimensional vector $\bs_k$ is the spreading code of the $k$-th user; we assume that the entries of $\bs_k$ are real and that $\bs_k^T \bs_k=\|\bs_k\|^2=1$, with $(\cdot)^T$ denoting transpose. Finally, $\bn$ is the ambient noise vector, which we assume to be a zero-mean white Gaussian random process with covariance matrix $(\N/2) \bI_N$, with $\bI_N$ the identity matrix of order $N$. An alternative and compact representation of (\ref{eq:r}) is given by
\beq
\br=\bS \bP^{1/2}\bH \bb + \bn \; ,
\label{eq:r2}
\eeq
wherein $\bS=[ \bs_1, \ldots, \bs_K]$ is the $N\times K$-dimensional spreading code matrix, $\bP$ and $\bH$ are $K \times K$-dimensional diagonal matrices, whose diagonals are $[p_1, \ldots, p_K]$ and $[h_1, \ldots, h_K]$, respectively, and, finally, $\bb=[b_1, \ldots, b_K]^T$ is the $K$-dimensional vector of the data symbols.

Assume now that each mobile terminal sends its data in packets of $M$ bits, and that it is interested both in having its data received with as small as possible error probability at the AP, and in making  careful use of the energy stored in its battery. Obviously, these are conflicting goals, since error-free reception may be achieved by increasing the received SNR, i.e. by increasing the transmit power, which of course comes at the expense of battery life\footnote{Of course there are many other strategies to lower the data error probability, such as for example the use of error correcting codes, diversity exploitation, and implementation of optimal reception techniques at the receiver. Here, however, we are mainly interested to energy efficient data transmission and power usage, so we consider only the effects of varying the transmit power, the receiver and the spreading code on energy efficiency.}. A useful approach to quantify these conflicting goals is to define the utility of the $k$-th user as the ratio of its throughput, defined as the number of information bits that are received with no error in unit time, to its transmit power \cite{nara1,nara2}, i.e.
\beq
u_k=\ds \frac{T_k}{p_k}\; .
\label{eq:utility}
\eeq
Note that $u_k$ is measured in bit/Joule, i.e. it represents the number of successful bit transmissions that can be made for each Joule of energy drained from the battery. The utility function (\ref{eq:utility}) is widely accepted and indeed it has been already used in a number of previous studies such as \cite{nara1,nara2,SaraydarPhD,meshkati,bacci,meshkati2}. Of course, there are also alternative choices that could be made. For instance, papers \cite{dey1,dey2} consider an outage-based utility suited for fast time-varying channels, while the recent study \cite{popescu} considers an utility that is the product of the transmit power times the interference\footnote{Obviously in this case we are interested to utility minimization rather than to its maximization.}. Utility (\ref{eq:utility}), however, is by no doubt the most suited one when energy efficiency is to be taken into account, and from now on we will definitely embrace this model.

Denoting by $R$ the common rate of the network (extension to the case in which each user transmits with its own rate $R_k$ is quite simple) and assuming that each packet of $M$ symbols contains $L$ information symbols and $M-L$ overhead symbols, reserved, e.g., for channel estimation and/or parity checks, the throughput $T_k$ can be expressed as
\beq
T_k=\ds R \frac{L}{M} E_k
\label{eq:Tk}
\eeq
wherein $E_k$ denotes the the probability that a packet from the $k$-th user is received error-free. In the considered DS/CDMA setting, the term $E_k$ depends formally on a number of parameters such as the spreading codes of all the users and the diagonal entries of the matrices $\bP$ and $\bH$, as well as on the strength of the used error correcting codes. However, a customary approach is to model the multiple access interference as a Gaussian random process, and assume that $E_k$ is an increasing function of the $k$-th user's Signal-to-Interference plus Noise-Ratio (SINR) $\gamma_k$, which is naturally the case in many practical situations.

Recall that, for the case in which a linear receiver is used to detect the data symbol $b_k$, according, i.e., to the decision rule
\beq
\widehat{b}_k=\mbox{sign}\left[\bd_k^T \br\right] \; ,
\label{eq:decrule}
\eeq
with $\widehat{b}_k$ the estimate of $b_k$ and $\bd_k$ the $N$-dimensional vector representing the receive filter for the user $k$, it is easily seen that the SINR $\gamma_k$ can be written as
\beq
\gamma_k=\ds \frac{p_k h_k^2 (\bd_k^T \bs_k)^2}{\frac{\N}{2}\|\bd_k\|^2 + \ds \sum_{i \neq k} p_i h_i^2
(\bd_k^T \bs_i)^2} \; .
\label{eq:gamma}
\eeq
Of related interest is also the mean square error (MSE) for the user $k$, which, for a linear receiver, is defined as
\beq
{\rm MSE}_k= E \left\{ \left(b_k - \bd_k^T \br \right)^2 \right\}=1 + \bd_k^T \bM \bd_k - 2\sqrt{p_k} h_k \bd_k^T
\bs_k \; ,
\label{eq:msek}
\eeq
wherein $E\left\{ \cdot \right\}$ denotes statistical expectation and $\bM=\left(\bS \bH \bP \bH^T \bS^T + \frac{\N}{2} \bI_N\right)$ is the covariance matrix of the data.

\medskip

The exact shape of $E_k(\gamma_k)$ depends  on factors such as the modulation and coding type.
However, in all cases of relevant interest, it is an increasing function of $\gamma_k$ with a sigmoidal shape, and converges to unity as $\gamma_k \rightarrow + \infty$; as an example, for  binary phase-shift-keying (BPSK) modulation coupled with no channel coding, it is easily shown that
\beq
E_k(\gamma_k)=\left[1-Q(\sqrt{2\gamma_k})\right]^M \; ,
\label{eq:psr}
\eeq
with $Q(\cdot)$ the complementary cumulative distribution function of a zero-mean Gaussian random variate with unit variance. A plot of  (\ref{eq:psr}) is shown in Fig. 1 for the case $M=100$.

It should be noted that substituting (\ref{eq:psr}) into (\ref{eq:Tk}), and, in turn, into (\ref{eq:utility}), leads to a strong incongruence. Indeed, for $p_k \rightarrow 0$, we have $\gamma_k \rightarrow 0$, {\em but} $E_k$ converges to a small but non-zero value (i.e. $2^{-M}$), thus implying that an unboundedly  large utility can be achieved by transmitting with zero power, i.e. not transmitting at all and making blind guesses at the receiver on what data were transmitted. To circumvent this problem, a customary approach \cite{nara2,meshkati} is to replace $E_k$ with an {\em efficiency function}, say $f_k(\gamma_k)$, whose behavior should approximate as close as possible that of $E_k$, except that for $\gamma_k \rightarrow 0$ it is required that $f_k(\gamma_k)= o(\gamma_k)$. The function
$f(\gamma_k)=(1-e^{-\gamma_k})^M$ is a widely accepted substitute for the true probability of correct packet reception, and in the following we will adopt this model\footnote{See Fig. 1 for a comparison between the
probability $E_k$ and the efficiency function.}.
This efficiency function is increasing and S-shaped, converges to unity as $\gamma_k$ approaches infinity, and has a continuous first order derivative. Note that we have omitted the subscript $``k''$, i.e. we have used the notation $f(\gamma_k)$ in place of $f_k(\gamma_k)$ since we assume that the efficiency function is the same for all the users.

Summing up, substituting (\ref{eq:Tk}) into (\ref{eq:utility}) and replacing the probability $E_k$ with the above defined efficiency function, we obtain the following expression for the $k$-th user's utility:
\beq
u_k=R \ds \frac{L}{M} \frac{f(\gamma_k)}{p_k} \; , \quad \forall k=1, \ldots, K \; .
\label{eq:utility2}
\eeq

Now, based on the utility definition (\ref{eq:utility2}), many interesting questions arise concerning how each user may maximize its utility, and how this maximization affects utilities achieved by other users. Likewise, it is natural to question what happens in a non-cooperative setting wherein each user autonomously and selfishly tries to maximize its own utility, with no care for other users utilities. In particular, in this latter situation, is the system able to reach an equilibrium wherein no user is interested in varying its parameters since each action it would take would lead to a decrease in its own utility? And, also, what is the price to be paid in terms of performance loss due to the selfish behavior (i.e., lack of cooperation) of the users?
Game theory  provides means to study these interactions and to provide some useful and insightful answers to these questions.

Initially, game theory was applied in this context mainly as a tool to study non-cooperative scenarios wherein mobile users are allowed to vary their transmit power only (see \cite{nara1,nara2,SaraydarPhD}, for example) to maximize utility, and where conventional matched filtering is used at the receiver. Recently, instead, in  \cite{meshkati} such an approach has been extended to the cross layer scenario in which each user may vary its power and its uplink {\em linear} receiver, i.e. the problem of joint linear multiuser detection optimization and power control for utility maximization has been tackled. In the following, we will go further by considering and analyzing the case of spreading code choice, power control and {\em linear} multiuser detector design for utility maximization.

\subsection{The proposed non-cooperative game}
Formally, the game $\G$ proposed here can be described as the triplet $\G=\left[\K, \left\{\S_k\right\}, \left\{u_k\right\} \right]$, wherein $\K= \left\{1, 2, \ldots, K\right\}$ is the set of active users participating in the game,
$u_k$ is the $k$-th user's utility defined in (\ref{eq:utility2}), and
\beq
\S_k=[0, P_{k,\max}] \times \R^N \times \R_1^N \; ,
\label{eq:strategy}
\eeq
is the set of possible actions (strategies) that user $k$ can take. It is seen that $\S_k$ is written as the Cartesian product of three different sets, and indeed $[0, P_{k, \max}]$ is the range of available transmit powers for the $k$-th user (note that $P_{k, \max}$ is the maximum allowed transmit power for user $k$), $\R^N$, with $\R$ the real line, defines the set of all possible linear receive filters, and, finally,
$$
\R_1^N= \left\{\bd \in \R^N \; : \; \bd^T \bd=1 \right\} \; ,
$$
defines the set of the allowed spreading codes\footnote{Here we assume that the spreading codes have real entries; the problem of utility maximization with reasonable complexity for the case of discrete-valued entries is a challenging issue that will be considered in the future.} for user $k$.

Summing up, the proposed non-cooperative game to be considered in the following can be cast as the following maximization problem
\beq
\ds \max_{\S_k} u_k = \max_{p_k, \bd_k, \bs_k} u_k(p_k, \bd_k, \bs_k ) \; , \quad
\forall k=1, \ldots, K \; .
\label{eq:game}
\eeq

\section{Preliminary results and concepts}
Before proceeding further, we review here some basic definitions on game theory, and provide some results on the relationship between SINR maximization and MSE minimization in multiuser systems. The content of this section will reveal useful in the sequel of the paper.

\subsection{Nash equilibria and Pareto optimality}
We give here the definition of {\em Nash equilibrium}. Let
$$
(s_1, s_2, \ldots, s_K) \in \S_1 \times \S_2 \times \ldots \S_K
$$
denote a certain strategy $K$-tuple for the active users. The point $(s_1, s_2, \ldots, s_K)$ is a Nash equilibrium if for any user $k$, we have
$$u_k(s_1, \ldots, s_k, \ldots, s_K) \ge u_k(s_1, \ldots, s_k^*, \ldots, s_K)\; ,
$$
$\forall s_k^* \neq s_k\, .$ Otherwise stated, at a Nash equilibrium, no user can {\em unilaterally} improve its own utility by taking a different strategy. A fast reading of this definition might lead to think that at Nash equilibrium users' utilities achieve their maximum values. Actually, this is not the case, since the existence of a Nash equilibrium point does not imply that no other strategy $K$-tuple does exist that can lead to an improvement of the utilities of some users while not decreasing the utilities of the remaining ones. These latter strategies are usually said to be Pareto-optimal \cite{gtbook}.
Otherwise stated, at a Nash equilibrium, each user, provided that the other users' strategies do not change, is not interested in changing its own strategy. However, if some sort of cooperation would be available, users might agree to simultaneously switch to a different strategy $K$-tuple, so as to improve the utility of some, if not all, active users, while not decreasing the utility of the remaining ones.

\subsection{SINR maximization and MSE minimization}
We are now ready to state our first result.

\noindent
 {\bf Proposition 1:}
{\em Given the linear decision rule (\ref{eq:decrule}) and the SINR expression in eq. (\ref{eq:gamma}), consider the non-cooperative game
\beq
\ds \max_{\bd_k, \bs_k} \gamma_k \; , \qquad \forall k=1, \ldots, K \; ,
\label{eq:gamesir}
\eeq
with the constraint $\|\bs_k\|=1$. This game admits a unique\footnote{Here and in the following uniqueness of the linear receive filter $\bd_k$ is meant up to a positive scaling factor. Uniqueness of the spreading codes is instead intended with respect to the set of eigenvalues of the matrix $\bS \bH \bP \bH^T \bS^T$.} Nash equilibrium point, which coincides with the unique global minimizer (with respect to spreading code choice and linear receiver choice) of the total MSE (TMSE) defined as
\beq
{\rm TMSE}= \ds \sum_{i=1}^K {\rm MSE}_i  \; .
\label{eq:tmse}
\eeq
Moreover, the Nash equilibrium point is also Pareto-optimal.}

\noindent
{\bf Proof:} This proof is partly based on results that are scattered in other papers; for the sake of conciseness and to avoid useless reproduction of already known material, we use these results citing their origin but without proving them again.

First of all, recall that among linear multiuser detectors, the MMSE receiver is the one that maximizes the SINR of each user \cite{wangpoor}, thus implying that, in order to maximize its own SINR, each user is to adopt a linear MMSE receiver, i.e. we have $\bd_k=\sqrt{p_k} h_k \bM^{-1}\bs_k$. Substituting this last relation in eq.  (\ref{eq:gamma}), and
using standard linear algebra techniques, it is easily shown that
\beq
\gamma_k=p_k h_k^2 \bs_k^T \left(\bM - p_k h_k^2 \bs_k \bs_k^T\right)^{-1} \bs_k \; .
\label{eq:gammamse}
\eeq
Given eq. (\ref{eq:gammamse}), it is seen that the $k$-th user SINR is maximized taking $\bs_k$ equal to the eigenvector corresponding to the minimal eigenvalue of the covariance matrix of the $k$-th user interference
$\left(\bM - p_k h_k^2 \bs_k \bs_k^T\right)$.
So far nothing guarantees that this strategy leads to a stable equilibrium point.
On the other hand, if a linear MMSE receiver is used,
the following relation is well-known to hold
\beq
{\rm MSE}_k = \ds  \frac{1}{1+ \gamma_k} \; ,
\label{eq:sirmsek}
\eeq
thus implying that SINR maximization for the generic $k$-th user is equivalent to minimization of its MSE. Moreover, exploiting the results contained in the Appendix I of \cite{honig}, it can be shown that, in the considered setting, individual MSE minimization is equivalent to minimization of the TMSE, defined in (\ref{eq:tmse}). Following \cite{ulukusyener,ensuring,rose}, letting $\bD=[\bd_1, \ldots, \bd_K]$ and denoting by $(\cdot)^+$ Moore-Penrose pseudoinversion, it can be shown that the TMSE admits a unique global optimum, and that the iterations
\beq
\begin{array}{lll}
\bd_i=\sqrt{p_i} h_i \left(\bS \bH \bP \bH^T \bS^T + \frac{\N}{2} \bI_N\right)^{-1}\!\! \bs_i   & \forall i=1, \ldots, K \\
\bs_i=\sqrt{p_i} h_i \left(p_i h_i^2 \bD \bD^T + \mu_i \bI_N \right)^{+} \bd_i  & \forall i=1, \ldots, K
\end{array}
\label{eq:iterazioni}
\eeq
admit as unique stable fixed points spreading code sets that are the global minimizer of the total MSE. In the above relations, $\mu_i$ should be set so that $\|\bs_i\|=1$, and a procedure  for  efficiently finding the value of $\mu_i$ for ensuring this constraint is given in Appendix A.

So far, we have shown that the non-cooperative game in (\ref{eq:gamesir}) can be solved by minimizing, through iterations (\ref{eq:iterazioni}), the total MSE, and that these iterations are guaranteed to converge to the unique and stable global optimum, i.e. the non-cooperative game admits a Nash equilibrium.
It remains to show the Pareto-optimality of this point. To this end, it suffices to show that, letting
 $\bar{\bS}$ and $\bar{\bD}$ be the spreading code matrix and the linear receiver matrix that jointly achieve the global minimum of the total MSE,  no strategy of spreading
codes and decoder can be found to increase the SINR of one or more users without decreasing the SINR of at least one other user.
To see this, note that if $\bar{\bS}$ and $\bar{\bD}$ are the global minimizers of the MSE, then $\bar{\bD}$ contains the MMSE receivers resulting from the spreading codes of $\bar{\bS}$. Denote by $\left\{ \gamma_i(\bar{\bS}, \bar{\bD})\right\}_{i=1}^K$ the SINR values achieved by the matrices $\bar{\bS}$ and $\bar{\bD}$. Assume now that there exists a spreading code matrix $\bS^* \neq \bar{\bS}$ such that $\gamma_i({\bS}^*, \bar{\bD}) > \gamma_i(\bar{\bS}, \bar{\bD})$, for at least one $i \in \{1, \ldots, K\}$ and
$\gamma_j({\bS}^*, \bar{\bD}) \geq \gamma_j(\bar{\bS}, \bar{\bD})$ for $j \neq i$. If this is the case, we can make an MMSE update and obtain the matrix $\bD^*$ of the MMSE receivers corresponding to the codes in $\bS^*$. For a given set of spreading codes, using the MMSE receiver always yields a maximization of the SINR and a minimization of the MSE. We thus have
$
\gamma_i(\bS^* , \bD^*)
 > \gamma_i(\bar{\bS}, \bar{\bD})$, and
$
\gamma_j(\bS^* , \bD^*)
 \geq \gamma_j(\bar{\bS}, \bar{\bD}) $, $\forall j \neq i$.
Consequently, given relation (\ref{eq:sirmsek}), we have
$$
{\rm TMSE}(\bS^* , \bD^*) < {\rm TMSE} (\bar{\bS}, \bar{\bD}) \; ,
$$
which contradicts the starting assumptions that $\bar{\bS}$ and $\bar{\bD}$ are the global minimizers of the MSE.
\hfill \rule{2mm}{2mm}

\section{A non-cooperative game for cross-layer resource allocation}
Equipped with the above result, we are now ready to resume the non-cooperative game in (\ref{eq:game}).
Note that, given  (\ref{eq:utility2}), the above maximization can be also written as
\beq
\ds \max_{p_k, \bd_k, \bs_k} \frac{f(\gamma_k(p_k, \bd_k, \bs_k))}{p_k} \; , \quad
\forall k=1, \ldots, K \; .
\eeq
Moreover, since the efficiency function is monotone and non-decreasing, we also have
\beq
\ds \max_{p_k, \bd_k, \bs_k} \frac{f(\gamma_k(p_k, \bd_k, \bs_k))}{p_k}= \max_{p_k}  \frac{f\left(\ds \max_{\bd_k, \bs_k}\gamma_k(p_k, \bd_k, \bs_k)\right)}{p_k} \; ,
\label{eq:deriv}
\eeq
i.e. we can first take care of SINR maximization with respect to spreading codes and linear receivers, and then focus on maximization of the resulting utility with respect to transmit power.

\medskip

We are now ready to express our result on the non-cooperative game for spreading code optimization, linear receiver design and power control.

\noindent
 {\bf Proposition 2:}
{\em The non-cooperative game defined in (\ref{eq:game}) admits a unique Nash equilibrium point $(p_k^*, \bd_k^*, \bs_k^*)$,
for $k=1, \ldots, K$, wherein
\begin{itemize}
\item[-]
$\bs^*_k$ and $\bd^*_k$ are the unique $k$-th user
spreading code and receive filter resulting from iterations (\ref{eq:iterazioni}). Denote by $\gamma_k^*$ the corresponding SINR.
\item[-]
$p_k^*=\min \{\bar{p}_k, P_{k, \max} \}$, with $\bar{p}_k$ the $k$-th user transmit power such that the $k$-th user maximum SINR $\gamma_k^*$ equals $\bar{\gamma}$, i.e. the unique solution of the equation $f(\gamma)=\gamma f'(\gamma)$, with $f'(\gamma)$ the derivative of $f(\gamma)$. \\
Moreover, for $K \leq N$, the Nash equilibrium point is Pareto-optimal.
\end{itemize}}
\noindent
{\bf Proof:} The proof generalizes the one provided in \cite{meshkati}, so, for the sake of brevity, we mainly focus on its original part.
Since $\partial \gamma_k /\partial p_k=\gamma_k /p_k$, it is easily seen that each user's utility is maximized if each user is able to achieve the SINR $\bar{\gamma}$, that is the unique\footnote{Uniqueness of $\bar{\gamma}$ is
ensured by the fact that the efficiency function is S-shaped \cite{rodriguez}.} solution of the equation $f(\gamma)=\gamma f'(\gamma)$. By Proposition 1, running iterations (\ref{eq:iterazioni}) until convergence is reached provides the set of spreading codes and MMSE receivers that maximize the SINRs for all the users. As a consequence, the utility of each user is maximized by adjusting transmit powers so that the optimized (with respect to spreading codes and linear receivers) SINRs equal $\bar{\gamma}$.
So far, we have shown how to  set the transmit power, spreading code and receiver design to maximize utility at the Nash equilibrium. In order to shown that a Nash equilibrium exists, we can use the same arguments of \cite{nara2} and state that a unique Nash equilibrium point exists since each user's utility function is quasi-concave\footnote{A function is quasi-concave if there exists a point below which the function is
nondecreasing, and above which the function is nonincreasing.} in the transmit power $p_k$ and since the efficiency function is S-shaped. \\
Assume now that $K \leq N$; in this case, the spreading codes resulting from iterations (\ref{eq:iterazioni}) are orthogonal, i.e. the multiuser channel boils down to $K$ parallel single-user channels. As a consequence, the SINR of each user is no longer affected by the strategies of the other users, and maximization of the utility of each user has no endangering effect on the utility achieved by the other users. In this scenario, thus, the non-cooperative game clearly achieves  Pareto optimality.
\hfill \rule{2mm}{2mm}

In practice, the above Nash equilibrium is reached through the following iterative algorithm. Given any set of transmit powers, iterations (\ref{eq:iterazioni}) are run in order to minimize system TMSE. After that, users adjust their transmit power in order to achieve the target SINR, using, e.g., the standard power control iterations as detailed in \cite{yates}. These steps are to be repeated until convergence is reached.

\begin{figure*}[!t]
\normalsize
\setcounter{MYtempeqncnt}{\value{equation}}
\setcounter{equation}{27}
\beq
\xi_i=\ds \frac{
\psi_iF^{-1}\left(\frac{K-i}{K}\right)
}
{
\N/2+ \ds \frac{K-u_2}{N}\frac{\psi_iF^{-1}\left(\frac{K-i}{K}\right) P_k}{\psi_iF^{-1}\left(\frac{K-i}{K}\right) +P_k \xi_i} + \frac{1}{N} \ds \sum_{j=K-u_2+1, j \neq i}^K
\frac{
\psi_iF^{-1}\left(\frac{K-i}{K}\right) P_{\max} F^{-1}\left(\frac{K-j}{K}\right)
}
{
\psi_iF^{-1}\left(\frac{K-i}{K}\right)+P_{\max}F^{-1}\left(\frac{K-j}{K}\right) \xi_i
}
} \; .
\label{eq:xi}
\eeq
\setcounter{equation}{\value{MYtempeqncnt}}
\hrulefill
\vspace*{4pt}
\end{figure*}

\section{A distributed power control algorithm based on LSA}

The above arguments show that implementation of the proposed non-cooperative game, as well as of the game in \cite{meshkati}, needs a power control algorithm, such as the one outlined in \cite{yates}.
Classical power control algorithms require knowledge of at least the uplink SINR for each user, or, alternatively, are implemented through iterative procedures \cite{yates2,stoc} that suffer from slow convergence and excess steady-state error. In this paper, instead, we show that LSA may lead to power control algorithms that may be implemented in a distributed fashion and that require knowledge of the channel for the user of interest only. The results of this section refer to the case that spreading code optimization is not performed, i.e. the case of utility maximization with respect to the choice of the linear uplink multiuser receiver and of the transmit power is considered.

As an introductory step to our algorithm, we begin by illustrating a simple power control algorithm derived from \cite{tse}.
We have mentioned that in a large CDMA system the $k$-th user's SINR converges in probability to the solution to Eq. (\ref{eq:HT}). Heuristically, this means that in a large system, and embracing the notation of the previous section, the SINR $\gamma_k$ is deterministic and approximately satisfies
\beq
\gamma_k\approx \ds\frac{h_k^2 p_k}{\N/2 + \ds \frac{1}{N} \sum_{j \neq k} \frac{h_k^2 h_j^2 p_k p_j}{h_k^2 p_k + h_j^2 p_j \gamma_k}}
\label{eq:HTheur}
\eeq
Now, as noted in \cite{tse}, if all the users must achieve the same common target SINR $\bar{\gamma}$, it is reasonable to assume that they are to be received with the same power, i.e. the condition
$$
h_1^2 p_1= h_2^2 p_2 = \ldots h_K^2 p_k = P_R\; ,
$$
is to be fulfilled. Substituting the above constraint in (\ref{eq:HTheur}) and equating (\ref{eq:HTheur}) to $\bar{\gamma}$ it is straightforward to come up with the following relation
\beq
P_R=\ds \frac{\bar{\gamma} \N/2}{1- \ds \frac{\gammabar}{1+\gammabar}\alpha}\; , \quad \Rightarrow \quad
p_k = \ds \frac{1}{h_k^2} \frac{\bar{\gamma} \N/2}{1- \ds \frac{\gammabar}{1+\gammabar}\alpha} \; ,
\label{eq:tsepc}
\eeq
wherein, we recall, $\alpha=K/N$, and the relation $\alpha< 1 + 1/ \bar{\gamma}$ must hold.
Eq. (\ref{eq:tsepc}), which derives from eq. (16) in \cite{tse}, gives a simple power control algorithm that permits setting the transmitted power for each user based on the knowledge of the channel gain for the user of interest only.
The above algorithm, however, does not take into account the situation which, due to fading and path losses, some users end up in transmitting at their maximum power without achieving the target SINR, and indeed our numerical results to be shown in the sequel will prove the inability of eq. (\ref{eq:tsepc}) to predict with good accuracy the actual power profile for the active users.

In order to circumvent this drawback, we first recall that in \cite{shamai} (see also \cite{husheng}) the following result has been shown:

\noindent
{\bf Lemma:} {\em Denoting by $F(\cdot)$ the cumulative distribution function (CDF) of the squared fading coefficients $h_i^2$, and  by $[h^2_{[1]}, \, h^2_{[2]}, \, \ldots, \, h^2_{[K]}]$ the vector of the users' squared fading coefficients sorted in non-increasing order, then we have that $h^2_{[\ell]}$ converges, for increasing number of users $K$, in probability to $F^{-1}\left(\frac{K-\ell}{K}\right)$, $\forall \ell=1, \ldots, K \, $.
}
\medskip

The above lemma states that if we sort a large number of identically distributed random variates, we obtain a vector that is approximately equal to the uniformly sampled version of the inverse of the common CDF of the random variates. Accordingly, in a large CDMA system each user may individually build a rough estimate of the fading coefficients in the network and be able to predict the number of users, say $u_2$,  that possibly will end up transmitting at the maximum power. Indeed, since, according to (\ref{eq:tsepc}) each user is to be received with a power $P_R$, the estimate $u_2$ of the number of users transmitting at the maximum power is given by
\beq
u_2= \ds \sum_{i=1}^{K}u \left(\ds \frac{\bar{\gamma} \N/2}{F^{-1}\left(\frac{K-i}{K}\right) \left(1- \alpha
\frac{\gammabar}{1+\gammabar}\right)} - P_{\max} \right) \; ,
\label{eq:u2}
\eeq
with $u(\cdot)$ the step-function. It is also obvious to assume that the users transmitting at $P_{\max}$ will be the ones with the smallest channel coefficients, i.e. the squared channel gains of the users transmitting at the maximum power are well approximated by the samples $F^{-1}\left(\frac{K-\ell}{K}\right)$, with $\ell=K-u_2+1, \ldots, K$. As a consequence, the generic $k$-th user will be affected by $u_1=K-u_2$ users that are received with power $P_R$ (these are the $u_1$ users with the strongest channel gains and that are able to achieve the target SINR $\gammabar$), and by $u_2$ users that are received with power $P_{\max} F^{-1}\left(\frac{K-\ell}{K}\right)$, with $\ell=K-u_2+1, \ldots, K$.
Denoting by $P_k$ the received power for the $k$-th user,
Eq. (\ref{eq:HTheur}) can be now written as
\beq
\begin{array}{ll}
\gamma_k= \\
\ds \frac{NP_k}{
\ds \frac{N\N}{2} + \frac{u_1P_kP_R}{P_k + P_R\gamma_k}+ \!\!\sum_{i=K-u_2+1}^K
\frac{P_k P_{\max} F^{-1}\left(\frac{K-i}{K}\right)}{P_k+ P_{\max} F^{-1}\left(\frac{K-i}{K}\right)\gamma_k}
}\; .
\label{eq:casino}
\end{array}
\eeq
Now, assuming for the moment that user $k$ is able to achieve its target SINR, i.e. that $P_k=P_R$, the second summand at the denominator on the RHS of the above equation can be approximated as
\beq
\frac{P_kP_R}{P_k + P_R\gamma_k}\approx \frac{P_k }{1+ \gamma_k} \; .
\eeq
Substituting the above approximations into (\ref{eq:casino}) and equating it to the target SINR $\gammabar$ we have
\beq
\ds \frac{ N P_k}{
\ds \frac{N\N}{2} + \frac{u_1 P_k}{1 + \gammabar}+ \sum_{i=K-u_2+1}^K
\frac{P_k P_{\max} F^{-1}\left(\frac{K-i}{K}\right)}{P_k+ P_{\max} F^{-1}\left(\frac{K-i}{K}\right)\gammabar}
} = \gammabar\; .
\label{eq:casino2}
\eeq
The above relation can be now numerically solved in order to determine the receive power $P_k$ for the  $k$-th user\footnote{Actually this equation gives the desired receive power for each user, and thus in a centralized power control algorithm needs to be solved just once.}; the actual transmit power for the $k$-th user is finally set according to the rule
\beq
p_k= \min \left\{P_k/h_k^2 , \, P_{\max} \right\}
\label{eq:pupdate}
\eeq

Summing up, the proposed algorithm may be summarized as follows. First, the number of users transmitting at the maximum power is estimated according to (\ref{eq:u2}). Then, the desired receive power for each user is computed solving eq. (\ref{eq:casino2}). Finally, the transmit power for the $k$-th user is determined according to relation (\ref{eq:pupdate}). Note that this algorithm requires knowledge only of the channel gain for the user of interest. In  Appendix B we will briefly sketch a method to compute the inverse of the CDF of the channel gains taking into account both fading and path losses due to random users' location with respect to the AP.

\section{Network performance prediction in a large CDMA system}
In this section we show how LSA arguments can be used to derive the utility, transmit power and achieved SINR profile across users in a large CDMA system. Otherwise stated, we show here that, based on the knowledge of the parameters $K$ and $N$, an estimate of the performance enjoyed by the ensemble of the users can be obtained.
We begin by considering the case in which no spreading code optimization is used, and, then, we will relax this constraint.

\subsection{Power control and linear MMSE detection}
Assume that no spreading code optimization is performed and that an MMSE linear multiuser detector is used at the receiver. Eq. (\ref{eq:pupdate}) provides the transmit power of the $k$-th user, wherein $P_k$ is the solution of eq. (\ref{eq:casino2}), and $h_k^2$ is the square of the channel coefficient for the $k$-th user. Once equation (\ref{eq:casino2}) has been solved (note that this equation is to be solved just once), the set $\cal P$ of the transmitted powers by the active terminals is expressed as
\beq
{\cal P}= \left\{ \min \left(P_k/F^{-1}\left(\frac{K-\ell}{K}\right) , \, P_{\max}\right)\right\}_{\ell=1}^K \; .
\label{eq:powerprofile}
\eeq
With regard to the set of the achieved SINRs, we have already commented on the fact that $K-u_2$ users are able to achieve the target SINR $\bar{\gamma}$. Denoting by $\xi_i$ the SINR achieved by the user whose channel coefficient is $h_{[i]}$, and letting $\psi_i$ denote the $i$-th element of the set $\cal P$ (i.e. $\psi_i=\min \left(P_k/F^{-1}\left(\frac{K-i}{K}\right) , \, P_{\max}\right)$), it is easily shown that $\xi_i$ can be approximately obtained as the solution to Eq. (28), shown at the top of this page.
\setcounter{equation}{28}
Accordingly, the set of the achieved SINRs in the network will contain $K-u_2$ elements equal to $\bar{\gamma}$ and $u_2$ elements given by the solution to the above equation with $i=K-u_2+1, \ldots, K$.
Given the set of transmit powers and of achieved SINRs, the set of achieved utilities will contain the elements
\beq
\ds
\upsilon_i=R\frac{L}{M} \frac{f(\xi_i)}{\psi_i} \; , \qquad i=1, \ldots, K \; .
\label{eq:upsilon}
\eeq

\subsection{Joint transmitter and receiver optimization with $K \leq N$}
Let us consider now the case that joint power control, spreading code optimization and linear receiver design is performed so as to maximize each user's utility. In this case, iterations (\ref{eq:iterazioni}) converge to a set of orthogonal codes, thus implying that the multiple-access channel boils down to the superposition of $K$ parallel single-user channels. In this case, the $k$-th user SINR $\gamma_k$ is expressed as
\beq
\gamma_k= \ds \frac{p_k h_k^2}{\N/2} \; , \quad k=1, \ldots, K\; .
\eeq
As a consequence, since each user should achieve a target SINR $\bar{\gamma}$, we have that the set $\cal P$ of the
transmitted powers is expressed as
\beq
{\cal P}= \left\{ \min\left(\ds \frac{\bar{\gamma} \N/2}{F^{-1}\left(\frac{K-i}{K}\right)}, P_{\max} \right)
\right\}_{i=1}^{K}
\label{P2}
\eeq
Accordingly, denoting by $\{ \xi_i\}_{i=1}^K$ the set of the achieved SINRs, we have
\beq
\xi_i=\ds \frac{\psi_i F^{-1}\left(\frac{K-i}{K}\right)}{\N/2} \; , \quad i=1, \ldots, K
\label{xi2}
\eeq
with $\psi_i=\min\left(\ds \frac{\bar{\gamma} \N/2}{F^{-1}\left(\frac{K-i}{K}\right)}, P_{\max} \right)$ the generic element of $\cal P$.
Given $\xi_i$ and $\psi_i$, the ensemble of achieved utilities can be computed as in (\ref{eq:upsilon}).

\subsection{Joint transmitter and receiver optimization with $K > N$}
Consider finally the case of an oversaturated CDMA system, i.e. the number of active users is larger than the processing gain. The following analysis refers to the case in which each user is able to achieve the target SINR $\bar{\gamma}$; it is thus reasonable to assume that all the users are received with the same power, i.e.
\beq
p_1h_1^2 = \ldots = p_K h_K^2=P_R \; .
\label{eq:equal}
\eeq
Since a linear MMSE detector is used at the receiver, using standard linear algebra it is easily shown that the $k$-th user SINR can be expressed as
\beq
\gamma_k=\bar{\gamma}= \ds \frac{p_k h_k^2 \bs_k^T \bM^{-1}\bs_k}{1- p_k h_k^2 \bs_k^T \bM^{-1}\bs_k}=
\ds \frac{P_R \bs_k^T \bM^{-1}\bs_k}{1- P_R\bs_k^T \bM^{-1}\bs_k} \; .
\label{eq:gamma5}
\eeq
On the other hand, it is well known \cite{ulukusyener,ensuring,optimal} that in the case in which $K>N$ and all the users are received with the same power, iterations (\ref{eq:iterazioni}) converge to a set of Welch-Bound-Equality (WBE) sequences, i.e. the limiting sequences are such that
\beq
\bS \bH \bP \bH^T \bS^T = P_R \alpha \bI_N    \; ,
\label{eq:wbe}
\eeq
where, we recall, $\alpha=K/N$. As a consequence, the data covariance matrix is expressed as
\beq
\bM=\left(P_R \alpha + \frac{\N}{2} \right)\bI_N \; .
\label{eq:covwbe}
\eeq
Substituting eq. (\ref{eq:covwbe}) into (\ref{eq:gamma5}) and solving for $P_R$ we have
\beq
P_R=\ds \frac{\bar{\gamma} \N/2}{1+ \bar{\gamma}(1-\alpha)} \; ,
\label{eq:PRR}
\eeq
with $\alpha< 1 +1/\bar{\gamma}$. Once $P_R$ has been computed from (\ref{eq:PRR}), the elements of the set of the transmitted powers are expressed as
\beq
\psi_i= \frac{P_R}{F^{-1}\left(\frac{K-i}{K}\right)} \; ,
\eeq
and the elements of the set of the achieved utilities can be computed as in (\ref{eq:upsilon}), with $\xi_i=\bar{\gamma}$, $\forall i=1, \ldots, K$.

\section{Socially optimum solution in the oversaturated scenario}
Proposition 2 has shown that the Nash equilibrium of the proposed non-cooperative game is also Pareto-optimal in the case in which $K\leq N$. For $K>N$, instead, the resource allocation strategy resulting from the said game is not on the Pareto-optimal frontier; the question thus arises on how much is the Nash equilibrium point far from the optimal frontier. Usually, the Pareto-optimal frontier cannot be easily computed, and an alternative and viable approach is to consider the following social problem
\beq
\max_{{\cal S}_1, \ldots, {\cal S}_K} \ds \sum_{i=1}^K u_i \; ,
\eeq
subject to the constraint of equal SINR, i.e. $\gamma_1= \ldots = \gamma_K= \gamma$, so that fairness among users can be ensured. The above problem can be thus written as
\beq
\max_{{\cal S}_1, \ldots, {\cal S}_K} \ds \sum_{i=1}^K u_i=
\max_{p_1, \ldots, p_K} \ds \sum_{i=1}^K \ds \frac{1}{p_i} \max_{\shortstack{$\bs_1, \ldots, \bs_k$ \\
$\bc_1, \ldots, \bc_k$}} f(\gamma)\; ,
\eeq
with $\gamma$ the common output SINR. Now, given the condition of equal SINR across users, it is natural to assume that the received powers are the same for all the users, i.e. eq. (\ref{eq:equal}) holds. Assuming that the optimal transmit power for all the users are smaller than $P_{\max}$, from eq. (\ref{eq:equal})  we have $p_i=P_R/h_i^2$, thus implying that the social problem becomes
\beq
\begin{array}{ccc}
\ds \max_{P_R} \frac{1}{P_R} \max_{\shortstack{$\bs_1, \ldots, \bs_k$ \\
$\bc_1, \ldots, \bc_k$}} f(\gamma) \ds \sum_{i=1}^K h_i^2 = \\
\ds \max_{P_R} \frac{1}{P_R} f \left( \max_{\shortstack{$\bs_1, \ldots, \bs_k$ \\
$\bc_1, \ldots, \bc_k$}} \gamma \right) \ds \sum_{i=1}^K h_i^2 \; .
\label{eq:u77}
\end{array}
\eeq
Since the received powers are the same for all the users, according to \cite{ulukusyener,optimal}, iterations (\ref{eq:iterazioni}) converge to a set of WBE sequences, which minimize the TMSE and, consequently, maximize the common SINR. Maximization of the common SINR with respect to the spreading codes and linear receivers of all the users can be thus carried out using iterations (\ref{eq:iterazioni}) after that the condition (\ref{eq:equal}) has been imposed. Let us denote by $\gamma^*$ the corresponding maximum common SINR; $\gamma^*$ is thus the SINR achieved by each user in a CDMA system wherein the spreading codes are WBE sequences, the receivers are MMSE detectors and each user is received with power $P_R$. Accordingly, we have
\beq
\gamma^*=  \ds \frac{P_R\bs_k^T \bM^{-1} \bs_k}{1-P_R\bs_k^T \bM^{-1} \bs_k} \; , \quad \forall k=1, \ldots, K
\eeq
Using eq. (\ref{eq:covwbe}), we obtain, after some algebra
\beq
P_R=\ds \frac{\gamma^* \N/2}{1- \gamma^*(\alpha-1)} \; , \quad \alpha<1+ \frac{1}{\gamma^*}
\eeq
Substituting the above relation into eq. (\ref{eq:u77}) the socially optimum problem is finally written as
\beq
\max_{\gamma} \ds \frac{f(\gamma)}{\gamma \N/2}\left[1- \gamma(\alpha-1)\right] \; .
\eeq
Taking the first order derivative of the above function with respect to $\gamma$ we have
\beq
\gamma f'(\gamma) \left[1- \gamma(\alpha-1)\right]=f(\gamma) \; .
\label{eq:gammasocial}
\eeq
The solution of this equation represents the utility-maximizing target SINR for each user in a social optimum context.

\section{Extensions to the Multi-Cell Scenario}
So far, we have considered the uplink of a single-cell scenario, i.e. out-of-cell interference has been either neglected or included in the additive thermal noise. However, real wireless networks are usually multi-cell, and users' utilities are affected also by the strategies of out-of-cell interference \cite{mandamulti}. In what follows, we thus give a brief look at the multi-cell case showing how the results of the previous sections may be extended to this case, pointing out some differences with the single-cell scenario, and revealing some interesting open issues for future investigations.

Let us thus consider the uplink of a multi-cell DS/CDMA wireless data network. Denote by $B$ the number of access points, and let $h_{i,j}$ be the real channel between the $j$-th user and the $i$-th AP; moreover, denote by $a(j)$ the index of the AP assigned to the $j$-th user\footnote{Note that we are assuming here that each user is assigned to a certain AP, i.e. AP assignments have already taken place.}. After chip-matched filtering and chip-rate sampling, the $N$-dimensional received data vector at the $\ell$-th AP, say $\br_\ell$, is written as
\beq
\br_\ell = \ds \sum_{k=1}^K \sqrt{p_k} h_{\ell, k} b_k \bs_k + \bn_\ell \; , \quad \ell=1, \ldots, B \; .
\label{eq:rell}
\eeq
The generic $k$-th user data is thus decoded at the $a(k)$-th AP, based on the decision rule
\beq
\widehat{b}_k=\sgn \left[\bd_k^T \br_{a(k)} \right] \; ,
\label{eq:decrule77}
\eeq
and the $k$-th user utility is now expressed as
\beq
u_k= R \ds \frac{L}{M} \ds \frac{f(\gamma_{a(k),k})}{p_k} \; ,
\label{eq:utility77}
\eeq
where, here, $\gamma_{a(k),k}$ is the $k$-th user SINR at the output of its linear receiver in its assigned AP, and is expressed as
\beq
\gamma_{a(k),k}=
\ds
\frac{p_k h^2_{a(k),k} (\bd_k^T \bs_k)^2}{\frac{\N}{2} \|\bd_k\|^2 + \ds \sum_{j\neq k}p_j h_{a(k),j}^2 (\bd_k^T \bs_j)^2
} \; .
\label{eq:gamma77}
\eeq
Of related interest is also the $k$-th user MSE achieved by the detection rule (\ref{eq:decrule77}); it is easy to show that it is expressed as
\beq
{\rm MSE}_k=1+ \bd_k^T \bM_{a(k)} \bd_k -2 \sqrt{p_k} h_{a(k),k} \bd_k^T \bs_k \; ,
\label{eq:mse77}
\eeq
with $\bM_{a(k)}= \ds \sum_{m=1}^K p_m h^2_{a(k),m} \bs_m \bs_m^T + \frac{\N}{2} \bI_N$ the covariance matrix of the data received that the $a(k)$-th AP.

Based on the above definitions, non-cooperative games for energy-efficient resource allocation can be considered in a multi-cell setting. Leaving aside for the moment the issue of spreading code allocation, the following result holds.

\noindent
{\bf Proposition 3:}
{\em Consider a non-cooperative game wherein the $k$-th user utility (\ref{eq:utility77}) is maximized with respect to the choice of the transmit power $p_k \in [0, P_{k,\max}]$ and of the linear receiver $\bd_k \in {\cal R}^N$. A unique Nash equilibrium point $(p_k^*, \bd_k^*)$
for $k=1, \ldots, K$, exists, wherein
\begin{itemize}
\item[-]
$\bd^*_k$ is the vector corresponding to a linear MMSE receiver;
\item[-]
$p_k^*=\min \{\bar{p}_k, P_{k, \max} \}$, with $\bar{p}_k$ the $k$-th user transmit power such that the $k$-th user  SINR $\gamma_{a(k),k}$ equals $\bar{\gamma}$, i.e. the unique solution of the equation $f(\gamma)=\gamma f'(\gamma)$, with $f'(\gamma)$ the derivative of $f(\gamma)$. \\
\end{itemize}}
\noindent
{\bf Proof:} The proof follows along the same lines of that of Proposition 2 and is omitted for the sake of brevity.
\hfill \rule{2mm}{2mm}

The above result states that, if transmit power and linear receiver are to be allocated, a unique Nash-equilibrium point does exist also in a multi-cell system. Unfortunately, things are more involved as optimization with respect to spreading codes too comes into play. If the total number $K$ of users in the network does not exceed the system processing gain $N$, then the following result holds.

\noindent
{\bf Proposition 4:}
{\em Consider a non-cooperative game wherein the $k$-th user utility (\ref{eq:utility77}) is maximized with respect to the choice of the transmit power $p_k \in [0, P_{k,\max}]$, of the linear receiver $\bd_k \in {\cal R}^N$ and of the spreading code $\bs_k \in {\cal R}_1^N$; assume that $K \leq N$.  A  Nash equilibrium point $(p_k^*, \bd_k^*, \bs_k^*)$
for $k=1, \ldots, K$, exists, wherein
\begin{itemize}
\item[-]
$\bs^*_k$ and $\bd^*_k$ are the $k$-th user
spreading code and receive filter minimizing the total MSE and can be obtained as fixed points of the iterations
\beq
\begin{array}{lll}
\bd_k=\sqrt{p_k} h_{a(k),k} \bM_{a(k)}^{-1} \bs_k \; , \quad & \forall k=1, \ldots, K \; , \\
\bs_k= \bd_k/\|\bd_k\| \; , \quad & \forall k=1, \ldots, K \; .
\label{eq:iterazioni77}
\end{array}
\eeq
Denote by $\gamma_k^*$ the corresponding SINR.
\item[-]
$p_k^*=\min \{\bar{p}_k, P_{k, \max} \}$, with $\bar{p}_k$ the $k$-th user transmit power such that the $k$-th user maximum SINR $\gamma_k^*$ equals $\bar{\gamma}$, i.e. the unique solution of the equation $f(\gamma)=\gamma f'(\gamma)$, with $f'(\gamma)$ the derivative of $f(\gamma)$.
\end{itemize}
This Nash equilibrium point is Pareto-optimal.
}\\

\noindent
{\bf Proof:} The proof is omitted for the sake of brevity.
\hfill \rule{2mm}{2mm}

Basically, the above result states that if $K\leq N$ a Nash equilibrium point does exist which is also Pareto-optimal; this point corresponds to the global minimum of the total MSE, which is a fixed point of iterations (\ref{eq:iterazioni77}). However, further investigation is needed to establish if other Nash equilibria may exist and, also, if iterations (\ref{eq:iterazioni77}) have some other fixed points corresponding to local minima of the total MSE. Likewise, the case in which $K>N$, which is the most relevant one in a multi-cell network, also merits some further investigation. These tasks are however  beyond the scope of this paper, and a thorough investigation of the multi-cell scenario, which is certainly worthwhile, is left for future work.

\section{Numerical Results}
In this section we illustrate some simulation results that give  insight into the performance of the proposed non-cooperative games, and, also, corroborate the validity of the analytical results of the previous sections.

We consider an uplink DS/CDMA system with processing gain $N=16$, and assume that the packet length is $M=120$;
for this value of $M$ the equation $f(\gamma)=\gamma f'(\gamma)$ can be shown to admit the solution $\bar{\gamma}=6.689 = 8.25$dB.
A single-cell system is considered, wherein users may have random positions with a distance from the AP ranging from 10m to 1000m. The channel coefficient $h_k$ for the generic $k$-th user is assumed to be Rayleigh distributed with mean equal to $d_k^{-1}$, with $d_k$ being the distance of user $k$ from the AP. We take the ambient noise level to be $\N=10^{-9}$W/Hz, while the maximum allowed power $P_{k,\max}$ is $-25$dBW. We present the results of averaging over $10000$ independent realizations for the users locations, fading channel coefficients and starting set of spreading codes. More precisely, for each iteration we randomly generate an $N \times K$-dimensional spreading code matrix with entries in the set $\left\{-1/\sqrt{N}, 1/\sqrt{N}\right\}$; this matrix is then used as the starting point for the games that include spreading code optimization, and as the spreading code matrix for the games that do not perform spreading code optimization.

Figs.  2 - 4 report the achieved average utility (measured in bits/Joule), the average user transmit power and the average achieved SINR at the receiver output versus the number of users, for the game in \cite{meshkati}, the game in \cite{nara2}  and for the non-cooperative game considered in Section IV of this paper. Inspecting the curves,
it is seen that the proposed approach largely outperforms the games of \cite{meshkati, nara2}. As an example,
it is seen that for $K=10$ users the utility achieved by the proposed game is about twice that achieved by the game in \cite{meshkati}, i.e. the same amount of energy can be used to transmit a doubled bulk of data.
In particular, it is seen that for $K\leq N$ a very substantial performance gain can be obtained by resorting to spreading code optimization; indeed, when $K\leq N$, users can be given orthogonal spreading codes, so that the multiaccess channel reduces to a superposition of $K$ separate single-user AWGN channels.
It is also seen from Fig. 4 that receivers achieve on the average an output SINR that is smaller than the target SINR $\bar{\gamma}$: indeed, due to fading and distance path losses, achieving the target SINR would require for some users a transmit power larger than the maximum allowed power $P_{k, \max}$, and so these users are not able to achieve the optimal target SINR. As a confirmation of this, in Fig. 5 we report the fraction of users transmitting at the maximum power: as expected, the smaller fraction corresponds to the proposed game, but it is seen that this fraction is larger than zero.

In order to validate the LSA-based distributed power control algorithm of Section V, we consider a system with processing gain $N=128$.
Fig. 6 reports the transmitted power profile across users for the  proposed distributed power control algorithm,
for the algorithm derived by Eq. (16) in \cite{tse} (i.e. eq. (\ref{eq:tsepc})), and for the conventional power control algorithm of \cite{yates}, that is non-adaptive and requires a substantial amount of prior information. It is seen that the proposed algorithm is capable of reproducing the optimal power profile with very good accuracy, while, on the contrary, the algorithm descending from paper \cite{tse} overestimates the required transmit powers and does not achieve a good performance.
While Fig. 6 shows the result of just one simulation trial (note however that a similar behavior has been observed in any considered case) the subsequent three figures report results coming from an average over 1000 independent realizations of the spreading codes, channel coefficients and users' locations. Figs. 7 - 9 show the achieved average utility (measured in bits/Joule), the average user transmit power and the average achieved SINR at the receiver output versus the number of active users, for the conventional power control algorithms (i.e. for the non-cooperative game of \cite{meshkati}), for the proposed algorithm, and for the power control algorithm derived by paper \cite{tse}. Results show that the proposed algorithm achieves a performance level practically indistinguishable from that of the standard algorithm, while the algorithm (\ref{eq:tsepc}) achieves an utility much smaller. From Fig. 9 it is however seen that the algorithm  (\ref{eq:tsepc}) achieves an output SINR larger than that of the other algorithms: this should not be interpreted as a sign of good performance. Indeed, in the considered scenario the aim of the power control algorithm is to make each user operate at a SINR equal to $\gammabar$.

Finally, we consider an oversaturated  system with processing gain $N=64$, and number of users $K=70$, so that $K>N$. In Fig. 10 we report the utility profile across users for the non-cooperative game proposed in Section IV, in comparison with the utility profile predicted according to the content of Section VI.C  and with the utility profile corresponding to the socially optimum solution with equal SINR constraint. It is seen that the performance loss incurred by the non-cooperative game in comparison with the socially optimum solution is quite negligible, and, also, that the LSA-based profile follows with good accuracy the actual utility profile. As a consequence, this plot corroborates the validity of our asymptotic analysis, that it is seen to be useful also when the system is actually ``not so large''.

\section{Conclusion}
In this paper the cross-layer issue of joint multiuser detection, power control, and spreading code optimization
for wireless data networks has been addressed. First of all, building on the study \cite{meshkati},  we have proposed a more general non-cooperative game wherein also spreading code optimization can be used to further increase the energy efficiency of CDMA-based wireless networks. We have shown that this game admits a unique Nash equilibrium point, that, for unsaturated systems, is also Pareto-optimal. For oversaturated CDMA systems, instead, we have shown that the socially optimum solution with equal SINR constraint exhibits a performance level practically coincident with that of the proposed non-cooperative game.
Using LSA, and assuming that no spreading code optimization is performed, a new distributed power control algorithm that can be implemented based on the knowledge of the channel for the user of interest only has been proposed. Additionally, through LSA results we have been able to derive the network utility profile for a large CDMA system, for both the cases that either spreading code optimization is carried out or it is not. Moreover, as an introductory step to the proposed non-cooperative game, we have clarified the relationship between the problems of SINR maximization and TMSE minimization in a synchronous CDMA system. Finally, we have also given a brief look at the multi-cell scenario, and, while extending some of our results to this case too, we have highlighted open issues worth being investigated in a future work.
Numerical results have confirmed the superiority of the proposed non-cooperative game with respect to competing alternatives, as well as that the LSA-based theoretical formulas describe with good accuracy the actual network performance.

\section*{Acknowledgments}
The authors wish to thank Dr. Husheng Li for insightful comments on a preliminary version of this paper. They are also grateful to the Associate Editor, prof. Lars Rasmussen, for his excellent management of the review process.

\section*{Appendix A}
Given the relation
$$
\bs_k=\sqrt{p_k} h_k \left(p_k h_k^2 \bD \bD^T + \mu_k \bI_N \right)^{+} \bd_k\; ,$$
we show here how to choose the constant $\mu_k$ so that $\|\bs_k\|=1$. Let $\bU \bLambda \bU^T$ be the eigendecomposition of the matrix $p_k h_k^2 \bD \bD^T$. Obviously, $\bU$ is an orthonormal matrix whose  columns are the eigenvectors of $p_k h_k^2 \bD \bD^T$, and $\bLambda$ is the corresponding diagonal eigenvalue matrix. Note that some of these eigenvalue will be zero for $K<N$. Now, letting
$\bu_i$ and $\lambda_i$ denote the $i$-th column  of $\bU$ and the $i$-th diagonal element of $\bLambda$, respectively, and
\beq
z(\lambda_i, \mu_k)=\left\{\begin{array}{llll} \frac{1}{\lambda_i + \mu_k} & \quad & {\rm if} \; \lambda_i + \mu_k \neq 0 \\
0 & \quad & {\rm if} \; \lambda_i+\mu_k=0 \; , \end{array} \right.
\eeq
it is easy to show that  the above spreading code update can be rewritten as
\beq
\bs_k=\sqrt{p_k} \ds \sum_{i=1}^{N} z(\lambda_i, \mu_k) \bu_i \bu_i^T \bd_k \; .
\label{eq:app2}
\eeq
From (\ref{eq:app2}) it is seen that, as $\mu_k\rightarrow + \infty$, $\|\bs_k\| \rightarrow 0$, thus implying that there exists a finite constant $Q_u$ such that $\|\bs_k\| < 1$ for any $\mu_k\geq Q_u$. Now, let $\lambda_m=\min_i{\lambda_i}$ (note that $\lambda_m$ may be 0 if $K<N$ or in general if $\bD \bD^T$ is not of full rank). It is easy to show that, as $\mu_k \rightarrow \lambda_m^+$, $\|\bs_k\|\rightarrow +\infty$. Accordingly, there exists a finite constant $Q_l >\lambda_m$ such that $\|\bs_k\| > 1$ for $\mu_k \in ]\lambda_m, Q_l]$. Since $\|\bs_k\|$ is monotonically decreasing for $\mu_k \in [Q_l, Q_u]$ and since $\|\bs_k\|>1$ for $\mu_k=Q_l$ and $\|\bs_k\|<1$ for $\mu_k=Q_u$, there exists just one value of $\mu_k$, say $\mu_k^*$, such that $\|\bs_k\|=1$ for $\mu_k=\mu_k^*$. The value of $\mu_k^*$ can be found using standard methods.

\section*{Appendix B}
In this appendix we show how the inverse CDF of the fading coefficients can be computed. In order to account for both fading and path loss, we assume that $h^2_k$ is given by the ratio of two random variables, i.e.
$
h^2_k= \ds \frac{\alpha_k^2}{d_k^{n}}$, wherein $\alpha_k$ is an exponential random variable (this corresponds to considering a Rayleigh fading channel), while $d_k$ is the distance of the $k$-th user from the BS; we assume that $d_k$ is uniformly distributed in the interval $[R_a,R_b]$; typical values may be $R_a=10m$ and $R_b=500m$. Finally $n$ is a non-random exponent; in urban environments $n$ is usually taken in the interval $[2, 5]$. It is easy to show that the CDF of $h_k^2$ is given by
\beq
F_{h_k^2}(x)= \Prob \left(h_k^2 \leq x\right) = E_{d_k} \left[1 - e^{-xd_k^n} \right]
\eeq
For the case $n=2$, straightforward computations lead to
\beq
F_{h_k^2}(x)=1 - \ds \frac{1}{R_b-R_a} \ds \sqrt{\frac{\pi}{x}}\left[\ds \frac{1}{2}\mbox{erfc}(R_a \sqrt{x})
- \frac{1}{2}\mbox{erfc}(R_b \sqrt{x}) \right] \; ,
\eeq
where erfc$(\cdot)$ is the complementary error function. The above equation should now be inverted numerically in order to obtain the inverse CDF. However, such an inversion may be computationally demanding, and, moreover, closed form expressions for the case $n \neq 2$ are not available.

An effective alternative approach is  the following. The interval $[R_a, R_b]$ can be partitioned in a given number, say $P$, of smaller intervals, and the probability density function of $d_k$ can be approximated as
\beq
f_{d_k}(d) \approx\ds \frac{1}{P} \sum_{i=1}^P \delta(d- R_a -i \Delta) \; , \quad \mbox{with} \; \; \; \Delta=\frac{R_b-R_a}{P}\; .
\eeq
As a consequence, we have
\beq
F_{h_k^2}(x)\approx 1- \ds \frac{1}{P} \sum_{i=1}^P e^{-xd_i^n} \; ,
\label{eq:CDFapprox}
\eeq
with $d_i=R_a + i \Delta$.
Equation (\ref{eq:CDFapprox}) is numerically invertible. Indeed, upon letting $e^{-x}=z$, we have
\beq
1- \ds \frac{1}{P} \sum_{i=1}^Pz^{d_i^n}= y \; \; \Rightarrow \; \; \ds  \sum_{i=1}^P z^{d_i^n}=(1-y)P \; .
\eeq
It is easy to see that the above equation admits a unique solution $z>0$, and any standard numerical equation solver
routine can be used to find it; after that, we have $x=-\ln(z)$, and this equals $F_{h_k^2}^{-1}(y)$.

In our simulations we have assumed $R_a=10m$, $R_b=1000m$ and $P=200$.


\begin{IEEEbiographynophoto}
{Stefano Buzzi}
(M'98 - SM '07) was born in Piano di Sorrento, Italy on December 10, 1970. He received with honors the Dr. Eng. degree  in 1994, and  the  Ph.D. degree in Electronic Engineering and Computer Science in 1999, both from the University of Naples "Federico II". In 1996 he spent  six months  at \emph{CSELT} (Centro Studi e Laboratori Telecomunicazioni), Turin, Italy, while he has had short-term visiting appointments at the Dept. of Electrical Engineering, \emph{Princeton University}, in 1999, 2000, 2001 and 2006. He is currently an Associate Professor at the University of Cassino, Italy. His current research and study interests lie in the area of statistical signal processing and resource allocation for wireless communications and radar applications.
Dr. Buzzi was awarded by the AEI (Associazione Elettrotecnica ed Elettronica Italiana) the "G. Oglietti" scholarship in 1996, and was the recipient of a NATO/CNR advanced fellowship in 1999 and of three CNR short-term mobility grants. He is currently serving as an Associate Editor for the \emph{IEEE Communications Letters}.
\end{IEEEbiographynophoto}

\begin{IEEEbiographynophoto}
{H. Vincent Poor}
 (S'72, M'77, SM'82, F'87) received
the Ph.D. degree in EECS from Princeton
University in 1977. From 1977 until 1990, he was
on the faculty of the University of Illinois at Urbana-
Champaign. Since 1990 he has been on the faculty
at Princeton, where he is the Michael Henry Strater
University Professor of Electrical Engineering and
Dean of the School of Engineering and Applied Science.
Dr. Poor's research interests are in the areas of
stochastic analysis, statistical signal processing and
their applications in wireless networks and related
fields. Among his publications in these areas is the recent book \emph{MIMO
Wireless Communications} (Cambridge University Press, 2007).
Dr. Poor is a member of the National Academy of Engineering, a Fellow
of the American Academy of Arts and Sciences, and a former Guggenheim
Fellow. He is also a Fellow of the Institute of Mathematical Statistics, the
Optical Society of America, and other organizations. In 1990, he served
as President of the IEEE Information Theory Society, and in 2004-2007 he
served as the Editor-in-Chief of the \emph{IEEE Transactions on Information Theory}.
Recent recognition of his work includes the 2005 IEEE Education Medal and
the 2007 IEEE Marconi Prize Paper Award.
\end{IEEEbiographynophoto}

\vfill


\newpage
\begin{onecolumn}

\begin{figure}
\centering
\includegraphics[width=10cm]{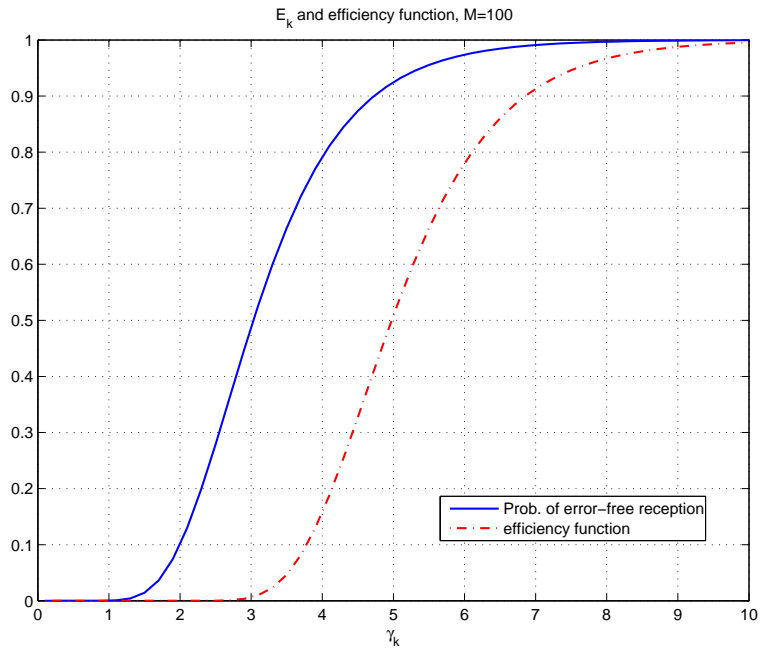}
\caption{Comparison of probability of error-free packet reception and efficiency function versus receive SINR and for packet size $M=100$. Note the S-shape of both functions.}
\label{figampiezza}
\end{figure}

\begin{figure}[t]
\centering
\includegraphics[width=10cm]{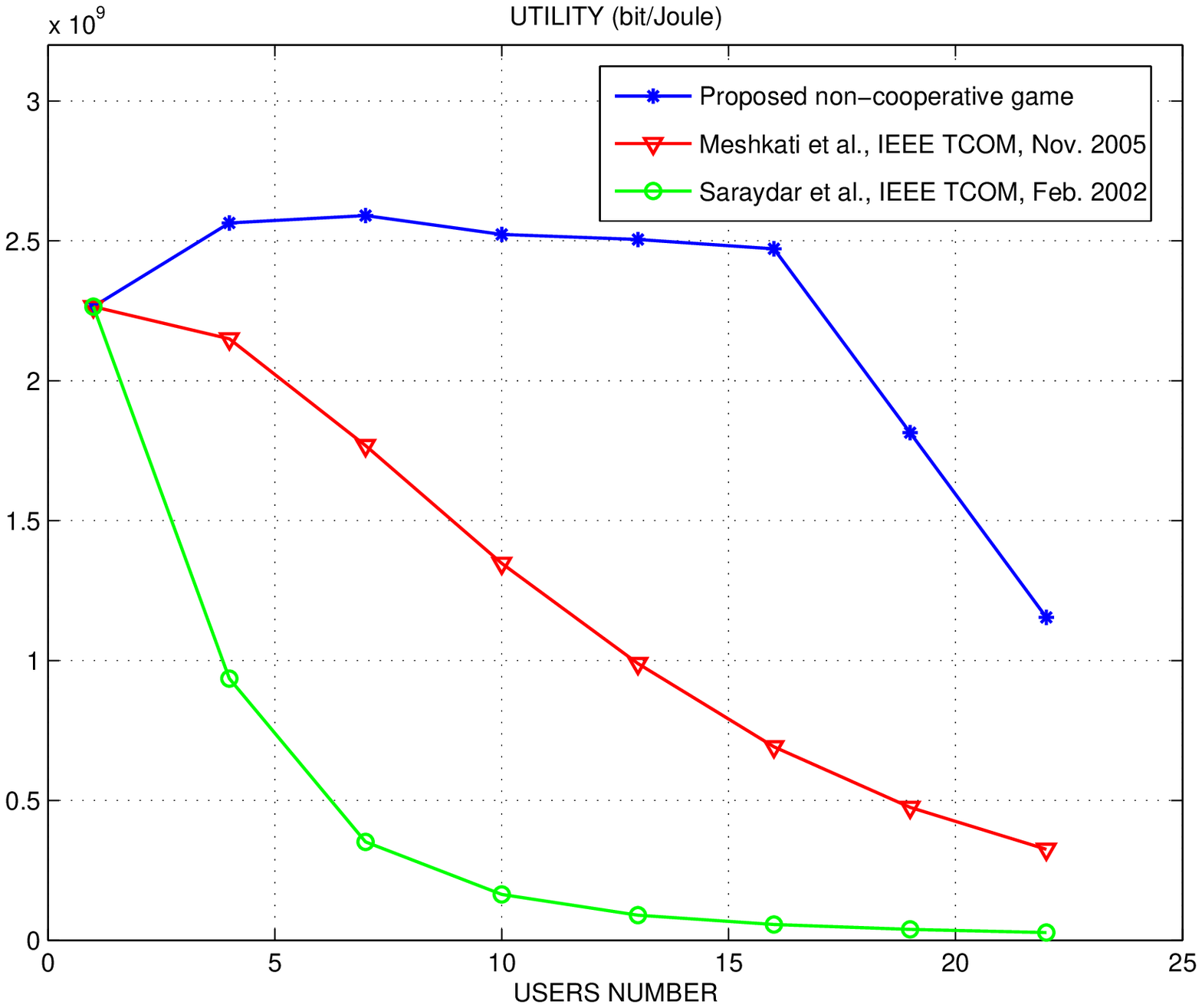}
\caption{Achieved average utility versus number of active users for the proposed noncooperative game and for the games in
references \cite{nara2} and \cite{meshkati}. The system processing gain is $N=15$.}
\end{figure}

\begin{figure}[t]
\centering
\includegraphics[width=10cm]{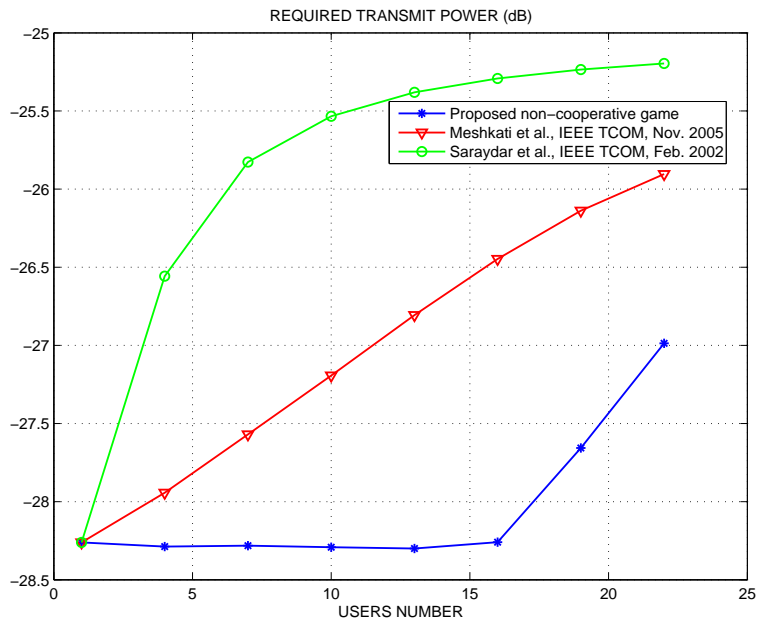}
\caption{Average transmit power versus number of active users for the proposed noncooperative game and for the game in references \cite{nara2} and \cite{meshkati}. The system processing gain is $N=15$.}
\end{figure}

\begin{figure}[t]
\centering
\includegraphics[width=10cm]{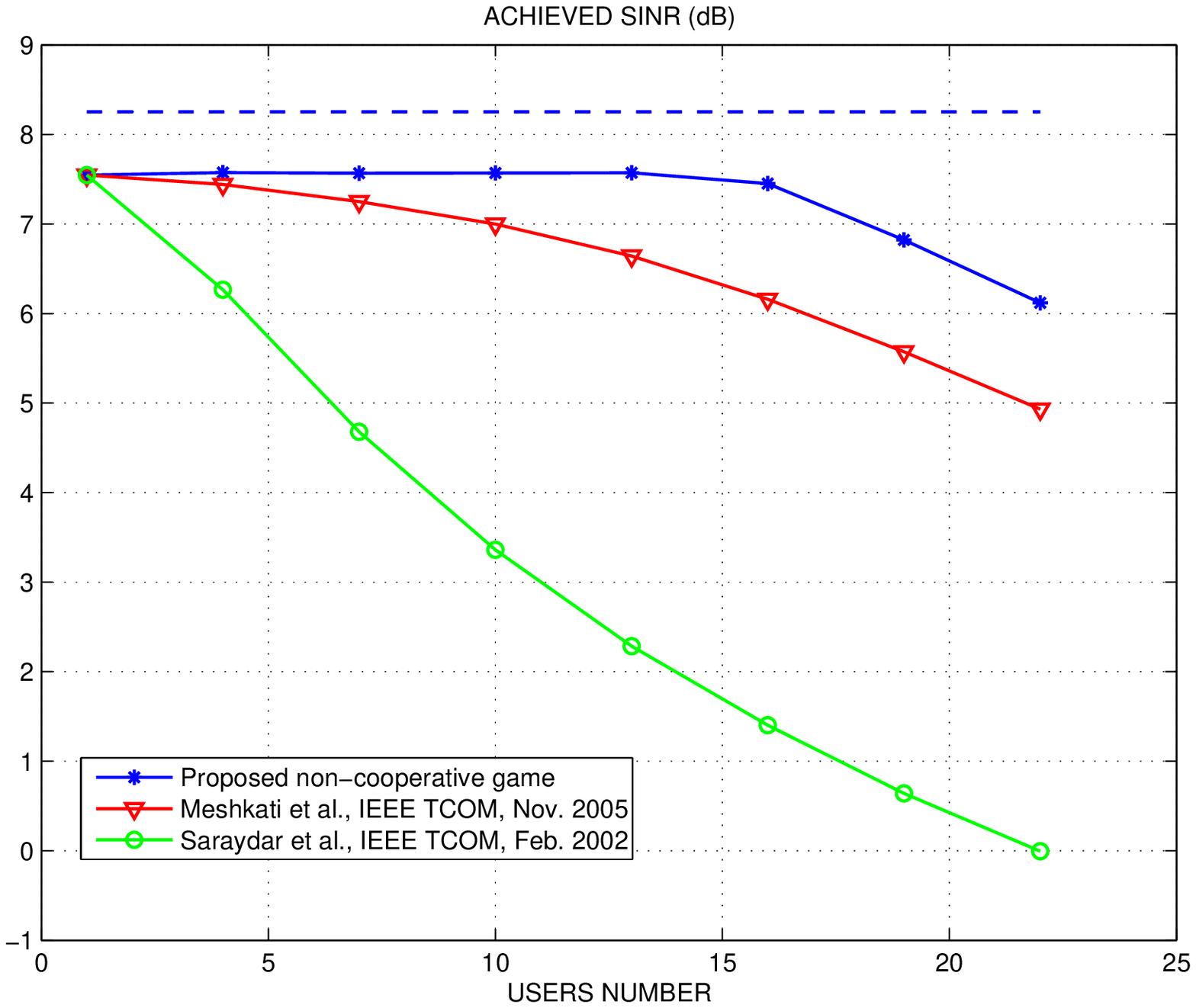}
\caption{Achieved average output SINR versus number of active users for the proposed noncooperative game and for the game in
references \cite{nara2} and \cite{meshkati}. The system processing gain is $N=15$.}
\end{figure}

\begin{figure}[t]
\centering
\includegraphics[width=10cm]{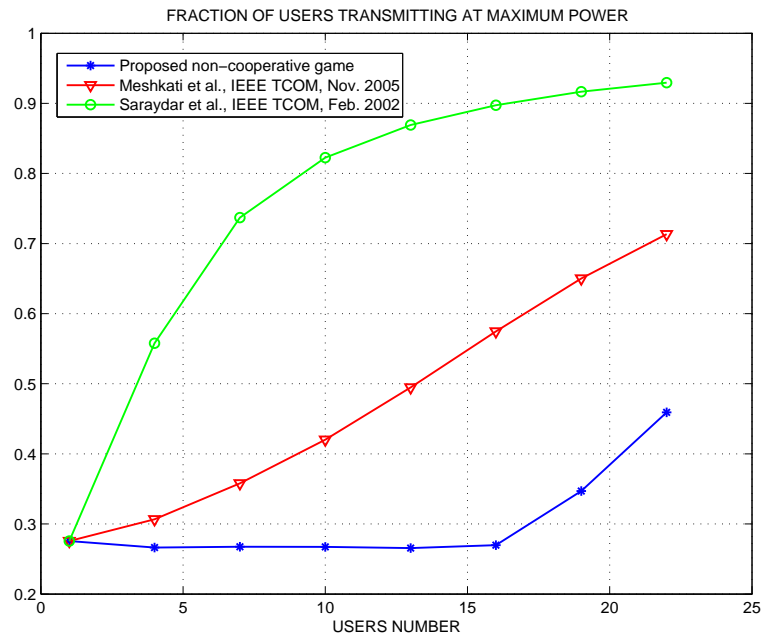}
\caption{Average fraction of users transmitting at their maximum allowed power versus number of active users for the proposed noncooperative game and for the game in
references \cite{nara2} and \cite{meshkati}. The system processing gain is $N=15$.}
\end{figure}

\begin{figure}
\centering
\includegraphics[width=10cm]{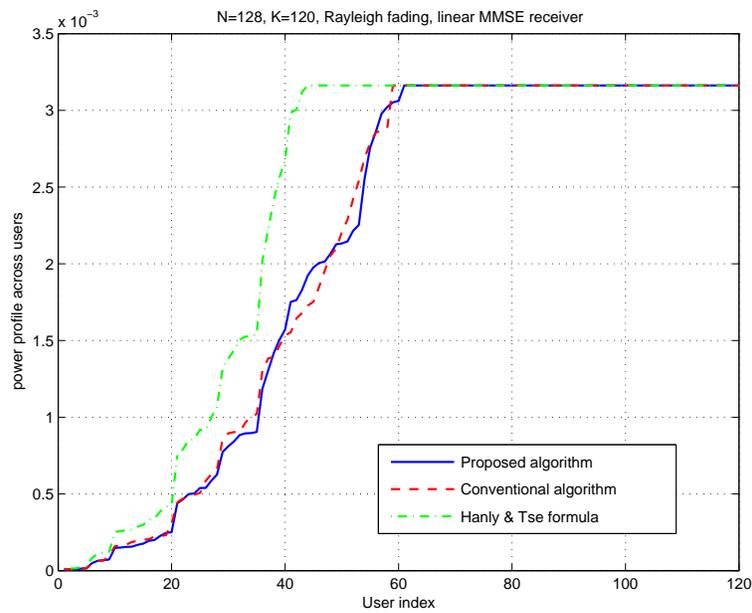}
\caption{Transmitted power profile across users for the proposed distributed algorithm based on LSA, the conventional power control algorithm \cite{yates} and the profile derived according to the algorithm in \cite{tse}.} \label{fig:9}
\end{figure}

\begin{figure}
\centering
\includegraphics[width=10cm]{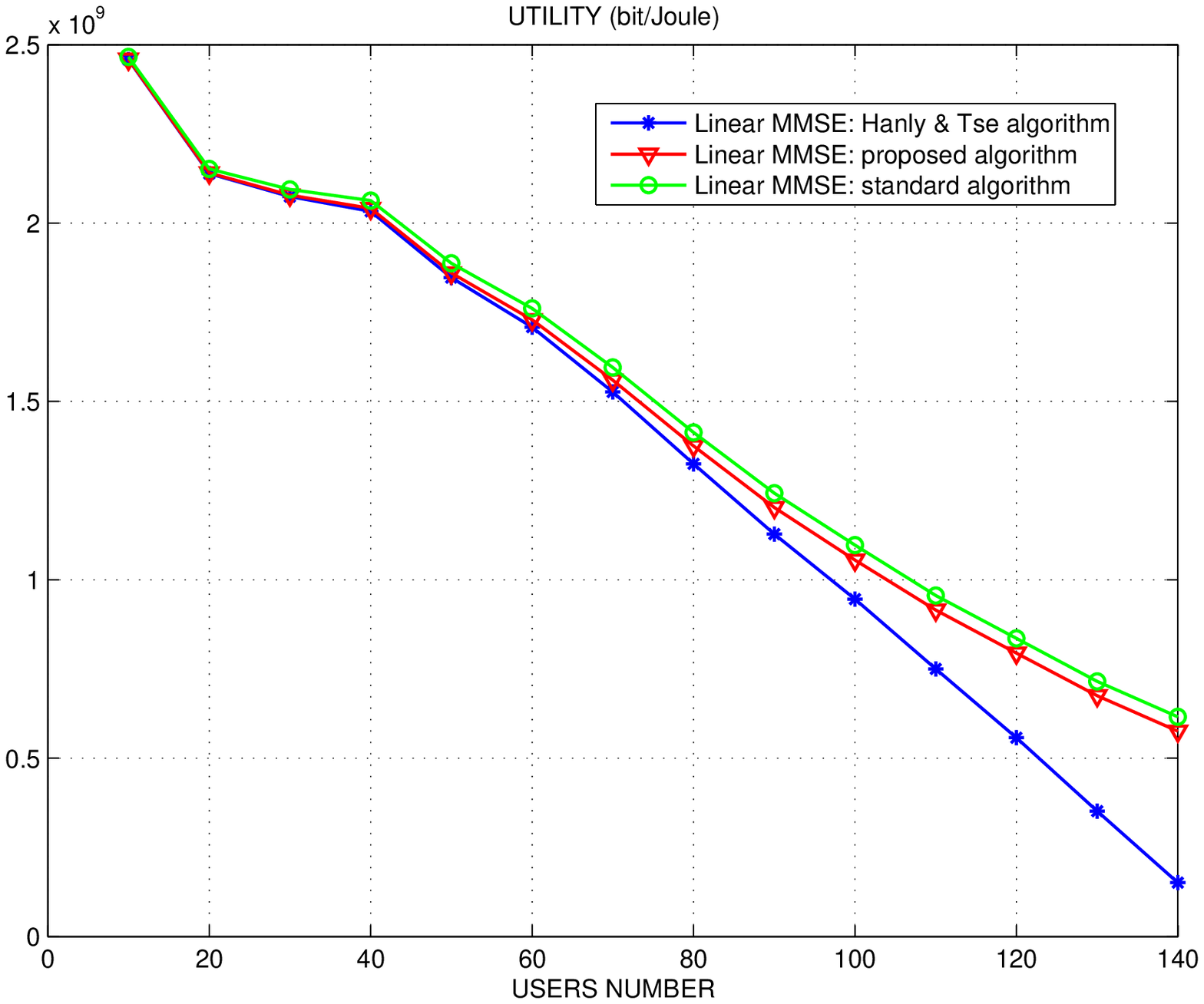}
\caption{Average utility versus number of users for the proposed distributed algorithm based on LSA, for the centralized implementation of reference \cite{meshkati} and for the distributed algorithm based on the power control algorithm of reference \cite{tse}.}
\end{figure}

\begin{figure}
\centering
\includegraphics[width=10cm]{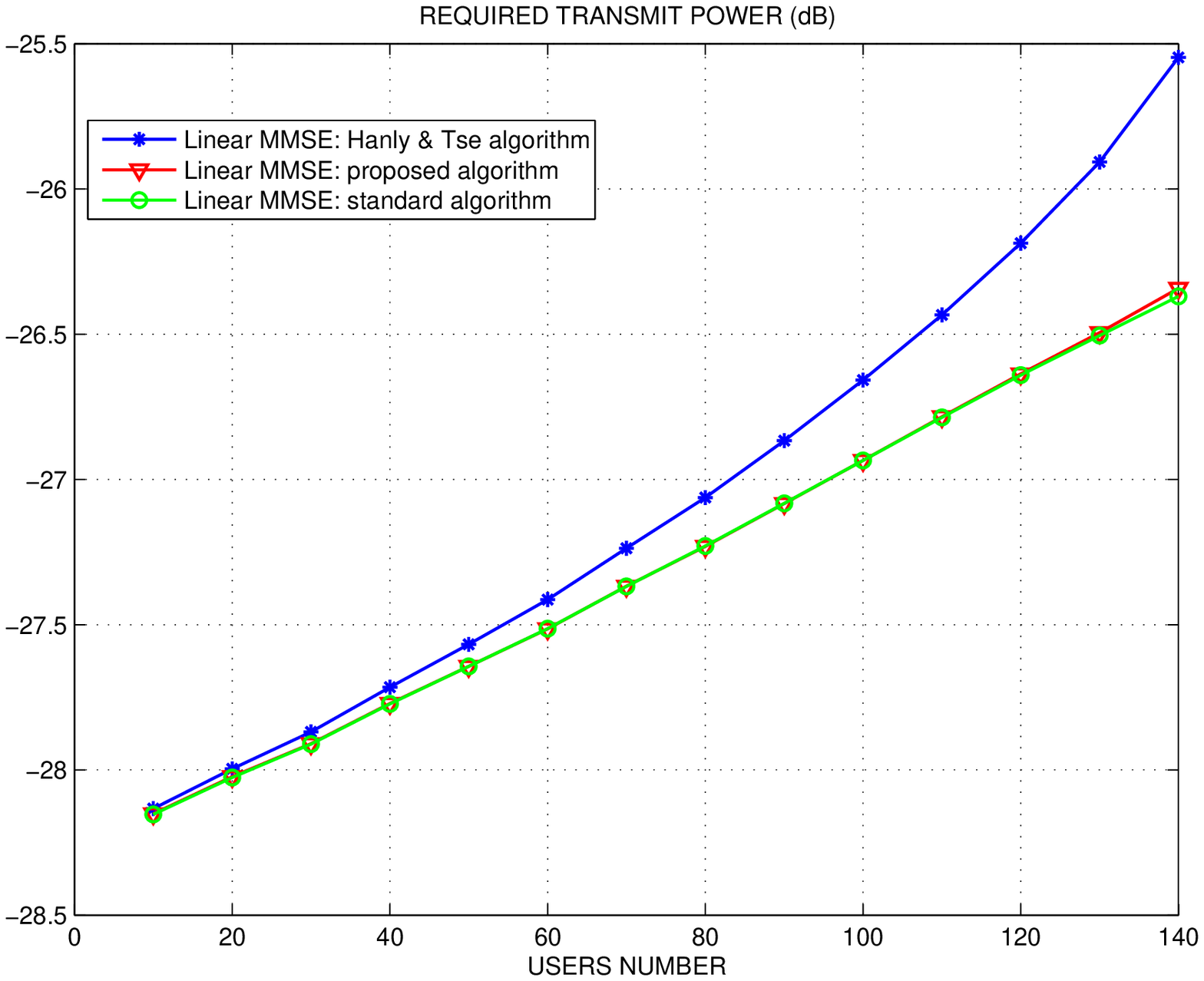}
\caption{Average transmit power versus number of users for the proposed distributed algorithm based on LSA, for the centralized implementation of reference \cite{meshkati} and for the distributed algorithm based on the power control algorithm of reference \cite{tse}.}
\end{figure}

\begin{figure}
\centering
\includegraphics[width=10cm]{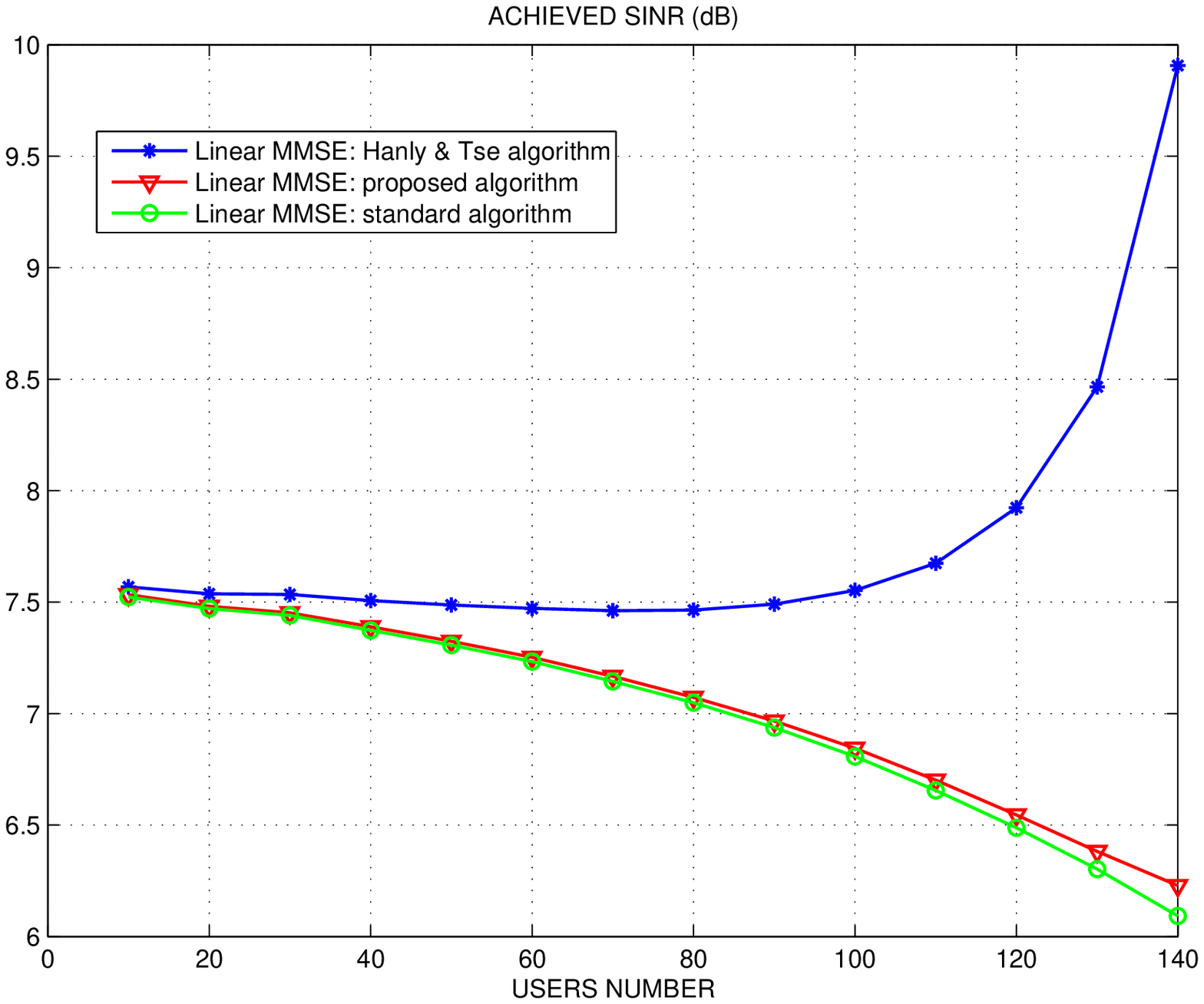}
\caption{Average achieved SINR versus number of users for the proposed distributed algorithm based on LSA, for the centralized implementation of reference \cite{meshkati} and for the distributed algorithm based on the power control algorithm of reference \cite{tse}.}
\end{figure}

\begin{figure}
\centering
\includegraphics[width=10cm]{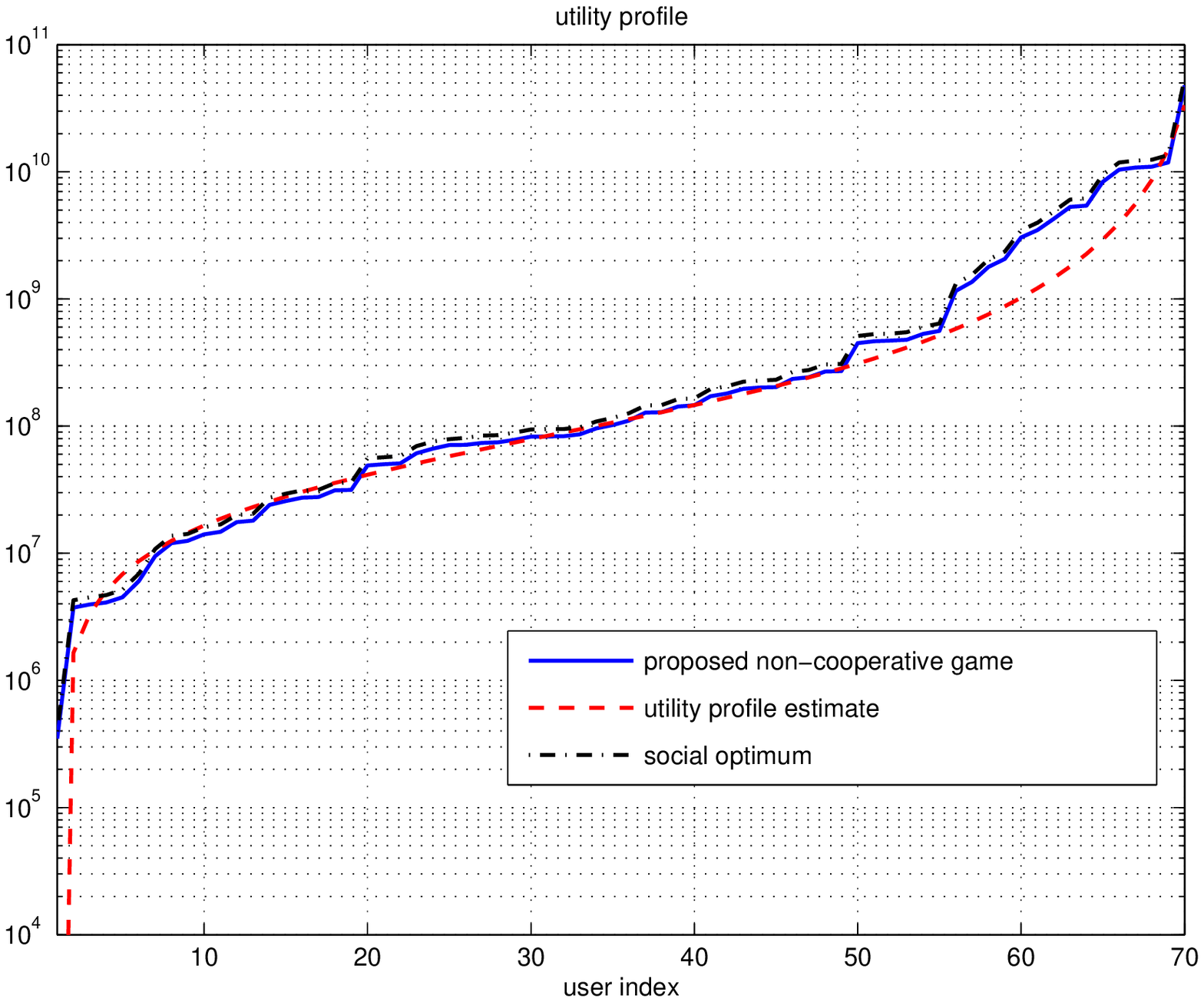}
\caption{Utility profile for the proposed non-cooperative game, in comparison with the social optimum
and with the utility profile predicted by large system analysis. Here the processing gain is $N=64$ and the number of users is $K=70$.}
\end{figure}

\end{onecolumn}

\begin{thebibliography}{99}

\bibitem{verdu}
S. Verd\'u, {\em Multiuser Detection}, Cambridge University Press, 1998.


\bibitem{wangpoor}
X. Wang and H. V. Poor,
{\em Wireless Communication Systems: Advanced Techniques for Signal Reception}. Upper Saddle River, NJ: Prentice-Hall, 2004.


\bibitem{crosslayer}
C. Comaniciu, N. B. Mandayam and H. V. Poor, {\em Wireless Networks: Multiuser Detection in Cross-Layer Design}, Springer, 2005.


\bibitem{gtbook}
D. Fudenberg and J. Tirole, {\em Game Theory}, Cambridge, MA: MIT Press, 1991.

\bibitem{gt}
A. B. MacKenzie and S. B. Wicker, ``Game theory in communications: Motivation, explanations, and applications to power control,''
{\em Proc. IEEE Global Telecommun. Conference}, San Antonio, TX, 2001.

\bibitem{yates}
R. D. Yates,
``A framework for uplink power control in cellular radio systems,''
{\em IEEE J. Sel. Areas Comm.}, Vol. 13, pp. 1341-1347, Sep. 1995.

\bibitem{nara1}
D. J. Goodman and N. B. Mandayam, ``Power control for wireless data,'' {\em IEEE Pers. Commun.}, vol. 7, pp. 48-54, Apr. 2000.

\bibitem{nara2}
C. U. Saraydar, N. B. Mandayam and D. J. Goodman, ``Efficient power control via pricing in wireless data networks,''
{\em IEEE Trans. Commun.}, vol. 50, pp. 291-303, Feb. 2002.

\bibitem{SaraydarPhD}
C. U. Saraydar, ``Pricing and power control in wireless data networks,''
Ph.D. dissertation, Dept. Elect. Comput. Eng., Rutgers University, Piscataway, NJ, 2001.

\bibitem{meshkati}
F. Meshkati, H. V. Poor, S. C. Schwartz and N. B. Mandayam,
``An energy-efficient approach to power control and receiver design in wireless data networks,''
{\em IEEE Trans. Comm.}, Vol. 53, pp. 1885-1894, Nov. 2005.

\bibitem{bacci}
G. Bacci, M. Luise, H. V. Poor and A. Tulino,
``Energy efficient power control in impulse radio UWB networks,''
{\em IEEE J. of Selected Topics in Sig. Proc.}, Vol. 1, pp. 508-520, Oct. 2007.


\bibitem{meshkati2}
F. Meshkati, H. V. Poor and S. C. Schwartz,
``Energy-efficient resource allocation in wireless networks: an overview of game-theoretic approaches,''
{\em IEEE Signal Proc. Magazine}, Vol. 24, pp. 58 - 68, May 2007.


\bibitem{tse}
D. N. C. Tse and S. V. Hanly,
``Linear multiuser receivers: effective interference, effective bandwidth and user capacity,''
{\em IEEE Trans. Inf. Theory}, Vol. 45, pp. 641-657, March 1999.

\bibitem{evans1}
L. G. F. Trichard, J. S. Evans, I. B. Collings,
``Large system analysis of linear multistage parallel interference cancellation,''
{\em IEEE Trans. Commun.}, Vol. 50, pp. 1778-1786, Nov. 2002.

\bibitem{evans2}
J. Evans and D. N. C. Tse,
``Large system performance of linear multiuser receivers in multipath fading channels,''
{\em IEEE Trans. Inf. Theory},
Vol. 46, pp. 2059-2078, Sept. 2000.


\bibitem{mimo}
J. Zhang and X. Wang,
``Large-system performance analysis of blind and group-blind multiuser receivers,''
{\em IEEE Trans. Inf. Theory},
Vol. 48, pp. 2507-2523, Sept. 2002.

\bibitem{dey1}
S. Dey and J. Evans,
``Optimal power control in wireless data networks with outage-based utility guarantees,''
{\em Proc. of the 42nd IEEE Conf. on Decision and Control}, Maui (HI), USA, Dec. 2003.

\bibitem{dey2}
T. Alpcan, T. Basar, and S. Dey,
``A power control game based on outage probabilities for multicell wireless data networks,''
{\em IEEE Trans. Wir. Commun.}, Vol. 5, pp. 890-899, April 2006.

\bibitem{popescu}
C. Lacatus and D. C. Popescu,
``Adaptive interference avoidance for dynamic wireless systems: a game theoretic approach,''
{\em IEEE J. of Selected Topics in Sig. Proc.}, Vol. 1, pp. 189-202, June 2007.



\bibitem{honig}
G. S. Rajappan and M. L. Honig, ``Signature sequence adaptation for DS/CDMA with multipath,''
{\em IEEE J. Sel. Areas Commun.}, Vol. 20, pp. 384-395, Feb. 2002.


\bibitem{ulukusyener}
S. Ulukus and A. Yener,
``Iterative transmitter and receiver optimization for CDMA networks,''
{\em IEEE Trans. Wireless Commun.}, Vol. 3, pp. 1879-1884, Nov. 2004.


\bibitem{ensuring}
P. Anigstein and V. Anantharam, ``Ensuring convergence of the MMSE iteration for interference avoidance to the global optimum,'' {\em IEEE Trans. Inform. Th.}, Vol. 46, pp. 873-885, Sept. 2000.

\bibitem{rose}
C. Rose,
``CDMA codeword optimization: Interference avoidance and convergence via class warfare,'' {\em IEEE Trans. Inf. Th.}, vol. 47, pp. 2368-2382, Sept. 2001.

\bibitem{rodriguez}
V. Rodriguez,
``An analytical foundation for resource management in wireless communication,''
{\em Proc. IEEE Global Telecommun. Conference}, San Francisco, CA, Dec. 2003.



\bibitem{yates2}
S. Ulukus and R. D. Yates,
``Stochastic power control for cellular radio systems,''
{\em IEEE Trans Commun.},
Vol. 46, pp. 784-798, June 1998.

\bibitem{stoc}
J. Luo, S. Ulukus and A. Ephremides,
``Standard and quasi-standard stochastic power control algorithms,''
{\em IEEE Trans. Inf. Theory},
Vol. 51, pp. 2612-2624, July 2005.


\bibitem{shamai}
S. Shamai (Shitz) and S. Verd\'u,
``Decoding only the strongest CDMA users,''
{\em Codes, Graphs and Systems}, R. Blahut and R. Koetter, Eds., pp. 217-228, Kluwer, 2002.



\bibitem{husheng}
H. Li and H. V. Poor,
``Power allocation and spectral efficiency of DS-CDMA systems in fading channels with fixed QoS-part I: single-rate case,''
{\em IEEE Trans. Wireless Commun.}, Vol. 5, pp. 2516-2528, September 2006.

\bibitem{optimal}
P. Viswanath and V. Anantharam, ``Optimal sequences and sum capacity of synchronous CDMA systems,''
{\em IEEE Trans. Inform. Th.}, Vol. 45, pp. 1984 - 1991, September 1999.
\newpage

\bibitem{mandamulti}
C. U. Saraydar, N. B. Mandayam and D. J. Goodman,
``Pricing and power control in a multicell wireless data network,''
{\em IEEE J. Sel. Areas Commun.}, Vol. 19, pp. 1883-1892, Oct. 2001.

\end{thebibliography}
\end{document}